\documentclass[reprint,showpacs,preprintnumbers,amsmath,amssymb, citeautoscript,prl,superscriptaddress]{revtex4-2}
\usepackage{graphicx,wasysym}
\usepackage{dcolumn}
\usepackage{bm,mathtools,booktabs}

\setcitestyle{super}

\usepackage[hypertexnames=false]{hyperref}

\usepackage{tikz}
\usetikzlibrary{positioning}
\usetikzlibrary{arrows.meta}
\usepackage{xcolor}
\usepackage{algorithm}
\usepackage{algpseudocode}

\makeatletter
\def\maketitle{
\@author@finish
\title@column\titleblock@produce
\suppressfloats[t]}
\makeatother

\begin{document}
\title{Navigation driven by bidirectional information transmission between sensing and actuation}

\preprint{APS/123-QED}
\author{Avishek Das}
\email{a.das@amolf.nl}
\affiliation{AMOLF, Science Park 104, 1098 XG, Amsterdam, The Netherlands\looseness=-1}
\author{Pieter Rein ten Wolde}
\email{p.t.wolde@amolf.nl}
\affiliation{AMOLF, Science Park 104, 1098 XG, Amsterdam, The Netherlands\looseness=-1}

\date{\today}
\begin{abstract}
A wide variety of biological functions are driven by feedback between sensing and actuation. A paradigmatic example is cellular navigation. 
During navigation, the
sensory system maps the environmental input signal onto a sensory output, which then drives an actuation response, thereby changing the future sensory input. How the accuracy of this bidirectional information transmission controls navigation is not currently understood. Here, we study how information controls navigation by analytically solving two generic models that describe two major classes of biological navigators: spatial- and temporal-sensing cells. We find that, in the linear-response regime of shallow gradients, navigation performance is fully determined by bidirectional information transmission alone, as quantified by feedforward and feedback transfer entropies. Elementary system parameters affect navigation performance only through their effect on these information flows. We call these relations \textit{Behavioral Equations of State} (BESTs): equalities that map information to function in a system-independent way. BESTs predict an experimentally testable data collapse for the performance of navigators with different sensing and actuation parameters.
We test the validity of our theory by performing stochastic simulations of chemotaxis of the bacterium \textit{Escherichia coli}, computing the relevant transfer entropies exactly with the TE-PWS algorithm. The observed performance obeys the BEST without any fitting or scaling parameters. Thus, our theory identifies  bidirectional information transmission between sensing and actuation as an organizing principle for navigation.
\end{abstract}
\maketitle

\thispagestyle{empty}
\twocolumngrid
Feedback between input and response is central to biological systems. Examples span scales: immune response to infection within an organism \cite{Ukogu.2026}, host-pathogen interactions during virus spreading through a population \cite{meijers2023population}, developing tissues that respond to morphogen gradients while simultaneously shaping them \cite{pezzotta2023optimal}, social interactions \cite{o2024dynamics}, and navigation. In all these cases, the input generates a response, the response reshapes the future input, and biological function emerges from this feedback between input sensing and actuation. Importantly, in biological systems sensing and actuation are typically stochastic. 
Information theory makes it possible to quantify their accuracy \cite{tkacik2025}, yet a  framework that can describe how navigation is enabled and constrained by information flow through sensing and actuation is currently lacking.

In this manuscript, we study this question using a minimal and experimentally accessible model system: cellular navigation. During navigation, the sensory system detects the environmental signal, relays it to the actuation system, which alters the cell’s swimming direction and thereby reshapes the future input.  Crucially, successful navigation requires accurate sensorimotor feedback: the sensory system must reliably encode the input into an internal representation, the sensory output, and the actuation system must reliably convert the sensory output into motion. To understand navigation, we must therefore quantify not only the information transmitted from the sensory input to the sensory output, but also the information transmitted from the 
sensory output, through the actuation system, back to the future sensory input. Equally important, we must also ask whether this bidirectional
information flow is functional: whether it actually improves navigation. However, while in the past decades the field has made tremendous progress in characterizing information transmission from input to output in the absence of feedback \cite{ziv2007optimal,mehta2009information,dubuis2011positional,tostevin2009mutual,cheong2011information,voliotis2014information,selimkhanov2014accurate,granados2018distributed,tjalma2023trade,mattingly2026coli}, and how that constrains navigation \cite{mattingly2021escherichia}, the flow of information in both directions, created by the sensorimotor feedback, has not been characterized so far. Moreover, how this bidirectional information flow enables and limits navigation has never been quantified for any stochastic navigator.

\begin{figure*}[t!]
\centering
\includegraphics[trim={2.2cm 0 0 0},clip,width=17.8cm]{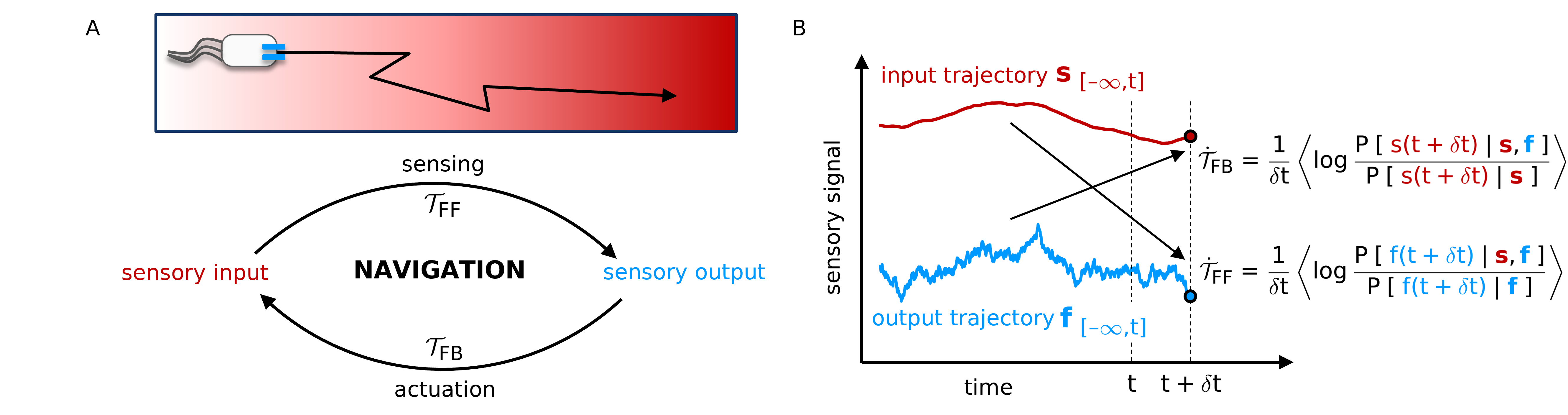}
\caption{Bidirectional information transmission drives navigation. (A) The chemoattractant concentration is the sensory input (red) for a navigating cell. The sensory system maps the input to a sensory output (blue). The reliability of this mapping is quantified by the feedforward information $\mathcal{T}_{\mathrm{FF}}$. The actuation system, in turn, responds to the sensory output, changing the cell's motion and thereby the future sensory input. This closes the feedback loop and gives rise to the feedback information transmission, $\mathcal{T}_{\mathrm{FB}}$. Navigation emerges from the interplay of $\mathcal{T}_{\rm{FF}}$ and $\mathcal{T}_{\rm{FB}}$. (B) Given steady-state trajectories of the sensory input and sensory output up to time $t$, $\mathbf{s}_{[-\infty,t]}$ and $\mathbf{f}_{[-\infty,t]}$ respectively, the feedforward transfer entropy rate $\dot{\mathcal{T}}_{\mathrm{FF}}$ is defined as the average log-likelihood ratio per unit time, with $\delta t \to 0,$ between predicting the future output from the past input and output trajectories, and predicting it from only the past output. The feedback transfer entropy rate $\dot{\mathcal{T}}_{\mathrm{FB}}$ is defined analogously, with the roles of $s(t)$ and $f(t)$ reversed.}
\label{fig:feedback}
\end{figure*}

Quantifying how navigation is controlled by bidirectional information flow is challenging. Both navigation and information are macroscopic observables that emerge from the interplay between the elementary parameters of the system, such as signaling gain, sensory noise, and actuation strength. To identify the relevant combinations of parameters that control navigation and information, analytical theory is vital. Moreover, the theory needs to recognize that in biological systems the information is often encoded not in the instantaneous values, but rather in the trajectories of the input and output signals \cite{tostevin2009mutual,selimkhanov2014accurate,granados2018distributed,marshall1995specificity,purvis2013encoding,levine2013functional}. For example, the bacterial chemotaxis system does not simply measure the current ligand concentration, but compares the current input with its recent past, thereby estimating a temporal derivative over its memory timescale \cite{segall1986temporal}. Last but not least, both the sensory and the actuation systems are often inherently nonlinear, and these nonlinearities tend to be amplified by the feedback between input sensing and actuation. 

To address these challenges, we start by analyzing minimal but generic models that can be solved analytically. Motivated by the observation that relatively large eukaryotic cells estimate a spatial derivative of the input concentration while relatively small and fast-moving bacteria take a temporal derivative \cite{berg1977physics}, we develop our theory for both spatial and temporal cellular sensing systems. The spatial-sensing model is linear and can be fully solved analytically using stochastic control theory, while the temporal-sensing model is weakly nonlinear, but can be solved by combining stochastic filtering theory with perturbation theory. Remarkably, for both modes of sensing, we find that in the linear-response regime of shallow gradients, navigation performance is determined by the strengths and timescales of only two information-theoretic quantities: the feedforward transfer entropy from sensory input to sensory output, and the feedback transfer entropy from sensory output, through actuation, back to the future sensory input. The system-specific parameters, like signaling gain, sensory noise and actuation strength, affect navigation only through their effects on information acquisition and transmission. This suggests that information acts as the universal currency for stochastic navigation across systems and scales. Importantly, this universal relation between navigation and information is obtained only when information is quantified via the multi-step transfer entropy, which accounts for information encoded in entire signal trajectories \cite{schreiber2000measuring}. Single-step transfer entropies are insufficient.
Strikingly, our analysis also reveals that generating information that is functional is non-trivial: when the actuation is too strong, both the feedforward and feedback information are high, yet the navigational performance is poor. 

Finally, to test whether our analytical predictions also hold beyond the minimal models, we study by simulations the chemotaxis system of the bacterium \textit{Escherichia coli}. This system is not only the best characterized sensorimotor system in biology, but it also provides a stringent test case because both the sensory pathway \cite{sourjik2002receptor} and the actuation system \cite{cluzel2000ultrasensitive}, the flagellar motor, are highly nonlinear. To compute the transfer entropies for this nonlinear system, we employ our recently developed TE-PWS algorithm, which makes it possible to quantify transfer entropies exactly for any stochastic model \cite{das2025exact}. Strikingly, we find that in the regime of shallow gradients, the relation between navigation and information as predicted by our analytical theory holds, without any fitting or scaling parameters. We refer to this equality as a Behavioral Equation of State (BEST): an equation that relates a macroscopic behavioral variable, navigation performance, to macroscopic information flows, without explicit reference to the microscopic parameters of the system.

\section*{Results}
To analytically investigate the role of information transmission in navigation, we construct two minimal dynamical models for sensing and actuation.
These models describe the two major navigation strategies employed by biological navigators: spatial and temporal sensing. Spatial-sensing cells estimate the chemoattractant gradient by comparing the concentration at the front and the back of the cell. This strategy is therefore used by navigators that are large enough for the difference in concentration across their cell length to be measurable, such as eukaryotic cells \cite{devreotes1988chemotaxis}. In contrast, smaller cells, such as prokaryotes, use temporal sensing: they infer the gradient from temporal changes in the attractant concentration  
while they swim \cite{berg1977physics}. 

Our models for spatial and temporal sensing are linear and weakly nonlinear, respectively. Each model incorporates all the key elementary parameters of generic navigators, such as sensory gain, adaptation time, and actuator gain. The most important feature of these models is, however, the presence of sensing and actuation noise.
This noise is physical in origin, arising from the stochasticity of the molecular sensing and actuation reactions \cite{berg1977physics,kaizu2014berg}. Such noise affects all general navigators with finite resource constraints, such as limited copy numbers of receptors and signaling molecules \cite{malaguti2021theory}, and a finite energy budget \cite{ouldridge2017thermodynamics}. 

Noise reduces the accuracy of both sensing and actuation. It therefore poses a severe challenge, because navigation can only be efficient if the sensory system constructs an accurate internal representation of the input signal, and if the actuation system accurately translates this internal representation into a new swimming direction. 
Therefore, navigation performance should depend on how accurately 
cells relay environmental information from the sensory input to the sensory output through the sensory system, and from the sensory output, through the actuation system, back to the future sensory input (see Fig. \ref{fig:feedback}A). These information flows can be quantified using information theory.

In information theory, the bidirectional information transmission between sensing and actuation in steady-state is quantified via transfer entropy rates \cite{schreiber2000measuring}.
Fig. \ref{fig:feedback}B illustrates the transfer entropy rate from the 
sensory input trajectory
$\mathbf{s}_{[-\infty,t]}$ to  the sensory output trajectory $\mathbf{f}_{[-\infty,t]}$. 
The transfer entropy rate is the average log-likelihood ratio between predicting the future output from the past input and output trajectories, and predicting it from the past output trajectory alone. It therefore quantifies how much the past input trajectory improves the prediction of the future output, beyond the information already contained in the past output trajectory itself.
We call this transfer entropy rate, which quantifies the rate of directional information transmission through the sensory system, the feedforward transfer entropy rate $\dot{\mathcal{T}}_{\mathrm{FF}}$. The transfer entropy rate from the sensory output, through the actuation system, back to the future input is defined analogously, and is called the feedback transfer entropy rate $\dot{\mathcal{T}}_{\mathrm{FB}}$.
Importantly, the transfer entropy rates account for information contained in the full past trajectories of the signals, and therefore effectively incorporate information stored in intermediate layers of signaling.

Nevertheless, accurate information transmission is not sufficient for efficient navigation. The transmitted information must also be functional: it must improve navigation. 
Biological navigators typically face different navigation objectives, depending on their environment, such as climbing gradients quickly, escaping from local concentration minima, or remaining near concentration
peaks \cite{celani2010bacterial,frankel2014adaptability}. To quantify navigation performance, we choose two paradigmatic navigation objectives in idealized environments: the drift speed up a linear concentration ramp, and the degree of localization near a harmonic concentration peak. For analytical tractability, we study the degree of localization in the spatial-sensing model and the drift speed in the temporal-sensing model.

\subsection*{Minimal model for spatial sensing}
\begin{figure*}[t!]
\centering
\includegraphics[trim={0 0 0 1.9cm},clip,width=17.8cm]{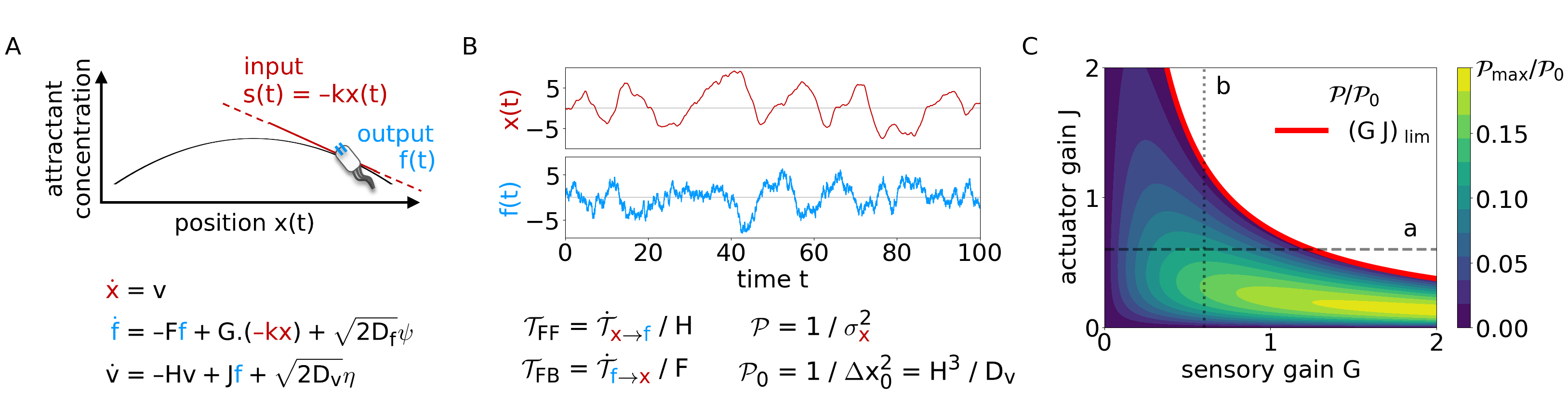}
\caption{Performance in spatial sensing depends on all elementary parameters. (A) Schematic illustration of the minimal model for spatial sensing. The attractant concentration profile is harmonic in position $x$ with curvature $k$. The sensory input is the gradient $-kx$, which is mapped to a sensory output $f$ with sensory gain $G$ and sensory noise $D_{f}$. The sensory output, in turn, modulates the velocity $v$ with an actuator gain $J$ and actuation noise $D_{v}$. The variables $f$ and $v$ relax to zero with rates $F$ and $H$, respectively. (B) Chemotaxis in this model localizes the cell near the concentration peak. Upper panels show a representative trajectory for $F=0.5,H=1,Gk=0.3,J=0.5,D_{f}=D_{v}=2$. 
Lower panels show the definitions of the feedforward and feedback information, and normalized performance, with $\Delta x_{0}$ denoting the persistence length of the cell in the absence of actuation. (C) Performance depends nontrivially on the elementary parameters $G,J,F$ and $H$. The color map shows the analytically obtained performance as a function of $G$ and $J$, with the other model parameters fixed at $F=0.5,H=1,D_{f}=D_{v}=2$. The red line denotes the stability boundary of the model, given by $GJ=FH(F+H)/k$. The dashed lines represent cross-sections at a fixed $J$ (line a) and a fixed $G$ (line b). Each line exhibits a maximum in performance at an intermediate gain. When both $G$ and $J$ are simultaneously optimized, the performance approaches a maximum $\mathcal{P}_{\mathrm{max}}$ that depends on the ratio $F/H$.}
\label{fig:spatial}
\end{figure*}

The minimal model for spatial sensing near a chemoattractant concentration peak is shown in Fig. \ref{fig:spatial}A. The position and the velocity of the cell are denoted by $x$ and $v$, respectively. The chemoattractant peak is quadratic in $x$ with a curvature $k$. The cell senses the gradient $-kx$ with a sensory gain $G$ and encodes its estimate of the gradient in the internal representation $f$. This sensory output then drives the actuation system  with an actuator gain $J$ to change the cell's velocity, which in turn affects its future position. Both $f$ and $v$ relax to their mean values, here taken to be zero, with rates $F$ and $H$, respectively. Additionally, both sensing and actuation are affected by Gaussian white noise, with respective noise magnitudes $D_{f}$ and $D_{v}$. 

The model exhibits chemotaxis through bidirectional feedback between noisy sensing and actuation. Fig. \ref{fig:spatial}B shows typical steady-state trajectories of $x(t)$ and $f(t)$. When the cell fluctuates to a large positive $x$, far away from the peak, the local gradient $-kx$ becomes strongly negative. The sensory system transmits this signal to the sensory output $f$, which therefore becomes anti-correlated with $x$. The negative sensory output then biases the velocity to negative values, driving the cell back to the peak. In this way, the bidirectional feedback keeps the cell localized near the peak.

We quantify navigation performance by this degree of peak localization. Specifically, as our performance metric $\mathcal{P}$, we choose the inverse steady-state variance of the cell's position $x$, $\mathcal{P} = 1 / \sigma^2_x$. In the localized steady-state, the bidirectional information transmission between sensing and actuation is quantified by the transfer entropy rates $\dot{\mathcal{T}}_{x\to f}$ and $\dot{\mathcal{T}}_{f\to x}$.

To compare performance with information, we first express both the navigation performance measure and the information metrics in dimensionless forms. We normalize the performance $\mathcal{P}$ by $\mathcal{P}_0 = 1/\Delta x_0^2$,  where  $\Delta x_0 = \sigma_{v,0}/H$ is the persistence length of the cell in the absence of actuation, i.e. the typical distance over which its velocity remains correlated.
Similarly, we normalize the feedforward and feedback transfer entropy rates by the memory timescales of the respective input signals in the absence of bidirectional coupling: $1/F$ for sensing and $1/H$ for actuation.

These observables are, in principle, experimentally measurable for a navigator. The position $x$ and velocity $v$ of the navigating cell can be measured through 3D tracking experiments \cite{taute2015high}. The internal representation $f$ can be measured by fluorescence microscopy, through the activity or concentration of signaling proteins \cite{sourjik2007vivo}. The transfer entropy rates can be quantified from the experimentally measured signal trajectories through the recently developed ML-PWS technique \cite{reinhardt2025ml}. The parameters $\sigma_{v0}$ and $H$ can be inferred from the velocity statistics of cells \textit{in which the coupling between sensing and actuation is removed}, for example using mutants \cite{korobkova2006hidden}. Similarly, $1/F$ can be measured from the dynamics of $f$\textit{in the absence of chemoattractant signals}. Therefore, this construction makes it possible to compute the dimensionless navigation performance,  feedforward and feedback transfer entropies, directly from experimental timeseries data.

\subsection*{Performance depends on all elementary parameters}
While our spatial-sensing model is minimal, it is still defined by six elementary parameters, all of which are system specific (Fig. \ref{fig:spatial}A). Which of these parameters control navigation performance? And can the performance be cast in a form that can be understood intuitively? To address these questions, we solve our model analytically (see Materials and Methods). This yields
\begin{align}
\mathcal{P}&=\frac{GkJ[FH(F+H)-GkJ]}{D_{f}(F+H)J^{2}+D_{v}[F^{2}(F+H)+GkJ]}.\label{eq:fullP}
\end{align}
Before analyzing the performance in terms of system specific parameters denoted by capital letters, we first describe how the performance depends on the steepness of the gradient $k$. In the limit of $k=0$, the performance goes to zero linearly. For intermediate values of $k$, as $k$ is increased, chemotaxis improves due to a stronger signal, but then decreases as the cell begins to overshoot the peak after each chemotactic climb. Finally, at $k_{\mathrm{lim}}=FH(F+H)/GJ$, the performance vanishes again. Beyond $k_{\mathrm{lim}}$, the linear model becomes unstable. In biological systems, such an instability would be avoided by bounds on the achievable values of $f$ and $v$, imposed by physical resource constraints \cite{govern2014optimal}.

We now turn to the question of how navigation performance depends on the elementary system parameters.
Eq. \ref{eq:fullP} shows that the performance depends on {\em all} the elementary parameters of the system, and in a highly non-trivial way. Fig. \ref{fig:spatial}C illustrates this non-trivial dependence, by showing how, for a given gradient $k$, performance varies with the sensory gain $G$ and actuation gain $J$, keeping the other system parameters constant. For any fixed value of $J$, performance is maximized at a nonzero value of $G$, and conversely, for any fixed value of $G$, performance is maximized at a nonzero value of $J$, as illustrated by the lines a and b. Crucially, these optimal gains themselves depend on the other elementary parameters. Thus, in the space of elementary parameters, sensing and actuation do not contribute independently to navigation performance. 

Eq. \ref{eq:fullP} further shows that parameters other than the sensory and actuation gains are also fundamental for navigation: they limit performance even after optimizing over $G$ and $J$. This optimization yields the maximal performance
\begin{align}
\frac{\mathcal{P}_{\mathrm{max}}}{\mathcal{P}_{0}}=\frac{F}{H}\left( 1+\frac{F}{H}\right) \left( \sqrt{1+\frac{F}{H}}-\sqrt{\frac{F}{H}}\right) ^{2}. \label{eq:maxP}
\end{align}
This expression shows that the relative speed of sensing over actuation, $F/H$, imposes an upper bound on navigation when all other elementary parameters can be optimized freely. In this regime, performance can be raised only by increasing the relative rate of sensory information processing, $F/H$, and $\mathcal{P}_{\mathrm{max}}$ increases monotonically with $F/H$.

Outside of this optimized regime, however, the behavior is far less trivial. In particular, at fixed values of $G$ and $J$, the performance is non-monotonic in $F$, decreasing as $1/F$ for large $F$. This optimum arises from the coupling between actuation noise and sensorimotor feedback: actuation noise, set by $D_v$, generates fluctuations in the sensory input, and a slow sensory response, i.e. small $F$, helps to time average this input noise \cite{becker2015optimal}. 

This analysis shows that the description in terms of elementary parameters is complete but not transparent. The dependence of navigation performance on the elementary parameters is highly coupled, and sensing and actuation cannot be optimized independently. We therefore ask whether navigation can instead be expressed in terms of a smaller set of collective variables that contribute more directly and independently to performance. We now turn to information theory to identify these variables.

\subsection*{Performance depends solely on bidirectional information transmission}
The natural candidates for these collective variables are 
the feedforward and feedback transfer entropies. We now test the central hypothesis of this paper, namely that, at least in a well-defined regime, navigation performance can be expressed in terms of these two information flows alone. 
To this end, we first compute the feedforward and feedback transfer entropies for the spatial-sensing model in terms of the elementary system parameters,
 using control theory \cite{chetrite2019information} (see Materials and Methods). The exact expressions are $\mathcal{T}_{\mathrm{FF}}=[\sqrt{1+(2Gk/H^2)(D_{v}/D_{f})^{1/2}}-1] /2$ and $\mathcal{T}_{\mathrm{FB}}=[ \sqrt{1+(J^{2}/F^2)(D_{f}/D_{v})}-1] /2$. We then combine these expressions with Eq. \ref{eq:fullP} to eliminate the elementary system parameters in favor of the two information-theoretic quantities. We find that in the regime of shallow gradients, i.e. small $k$, this gives the following equality:
\begin{align}
\frac{\mathcal{P}}{\mathcal{P}_{0}}=\frac{\mathcal{T}_{\mathrm{FF}}}{\frac{1}{2}\left[ \sqrt{(2\mathcal{T}_{\mathrm{FB}}+1)^{2}-1}+\frac{1}{\sqrt{(2\mathcal{T}_{\mathrm{FB}}+1)^{2}-1}}\right]. }\label{eq:smallkinfo}
\end{align}
This is the first central result of this paper. It shows that in the linear-response regime of shallow gradients navigation is controlled by bidirectional information flow alone. The elementary system parameters affect navigation only through their effect on information transmission. Moreover, the equality reveals that in this regime the performance factorizes into two independent contributions, from the feedforward and feedback transfer entropies, respectively. We refer to Eq. \ref{eq:smallkinfo} as a Behavioral Equation of State, because it relates navigation performance directly to the feedforward and feedback information flows, without explicit reference to the elementary system parameters.

The dependence of the navigation performance on information, Eq. \ref{eq:smallkinfo}, reveals deeper insights into navigation in shallow gradients. The performance increases monotonically with the feedforward information $\mathcal{T}_{\mathrm{FF}}$. In shallow gradients where sensory information is a limiting resource, more feedforward information is always useful. 
In contrast, the dependence on feedback information is non-monotonic:
performance is maximized at an optimal feedback information $\mathcal{T}^{*}_{\mathrm{FB}}=(\sqrt{2}-1)/2~\mathrm{nats}$. 
Too much feedback information is detrimental to performance.

\begin{figure}[t!]
\centering
\includegraphics[trim={0 0 0 0cm},clip,width=8.7cm]{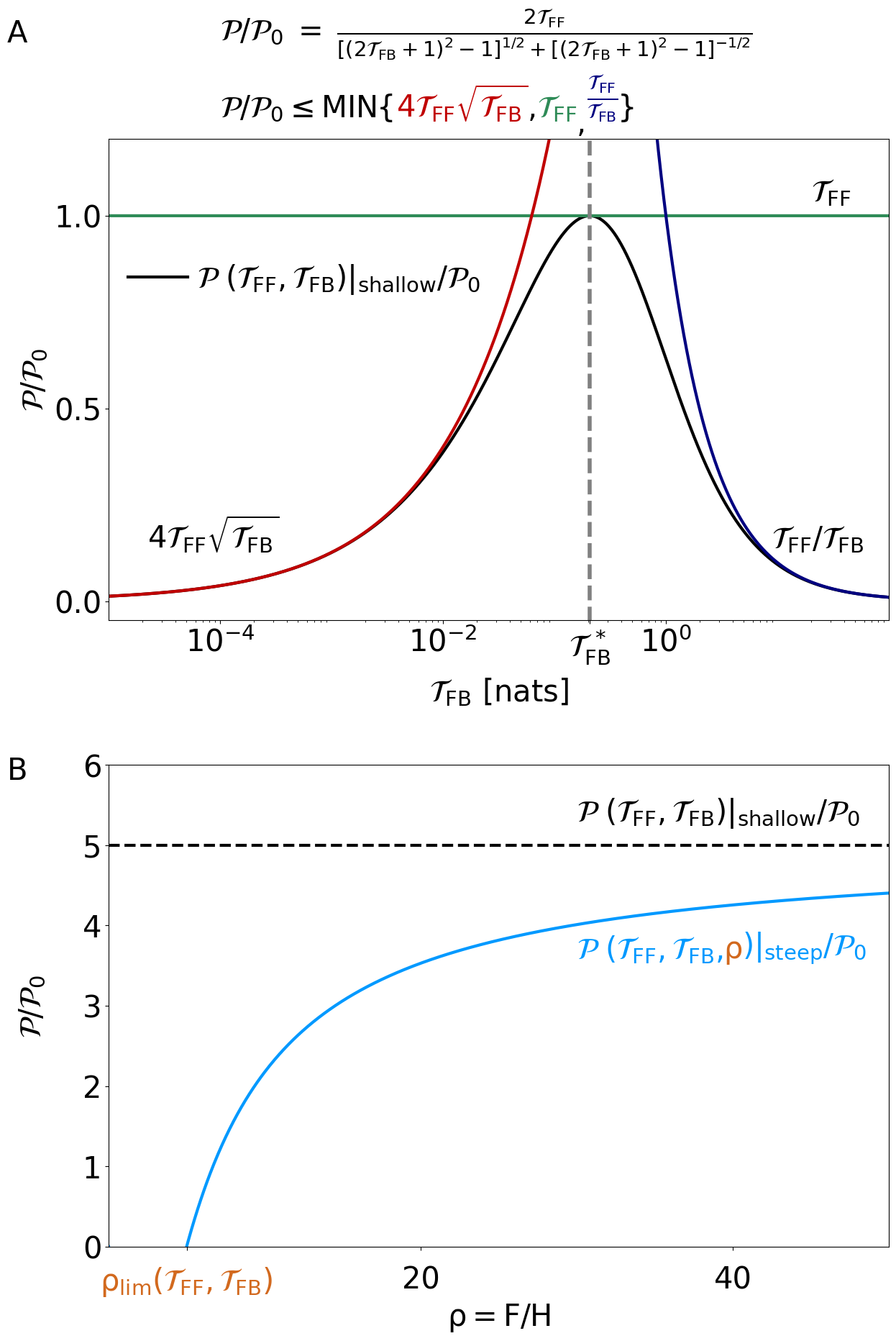}
\caption{In the spatial-sensing model, navigation performance is controlled by bidirectional information transmission.
 (A) In shallow gradients, navigation performance is controlled by the feedforward and feedback information alone. The black solid line illustrates the equality of  Eq. \ref{eq:smallkinfo}, showing how performance varies with feedback information at fixed feedforward information; for plotting, $\mathcal{T}_{\mathrm{FF}}$ has been set to $1~\rm{nat}$. The dashed grey line denotes the optimal feedback information. The colored solid lines indicate the three bounds on performance in the corresponding limiting regimes (Eq. \ref{eq:smallkbound}). (B) In steep gradients, the relative rate of  sensory information processing over actuation imposes an additional constraint on navigation, beyond the feedforward and feedback information. The panel shows how performance monotonically improves with $\rho=F/H$ at fixed feedforward and feedback information.  The linear model becomes unstable when $\rho<\rho_{\mathrm{lim}}=[(2\mathcal{T}_{\mathrm{FF}}+1)^{2}-1]\sqrt{(2\mathcal{T}_{\rm{FB}}+1)^{2}-1} \; /2 -1$. At large $\rho$, the performance asymptotically approaches the performance in shallow gradients for the same amounts of feedforward and feedback information. This curve was obtained by first computing the feedforward and feedback information for  $F=0.5,H=J=1,Gk=3,D_{f}=0.1,D_{v}=2$, and then varying $\rho$. }
\label{fig:spatialinfo}
\end{figure}

To elucidate the origin of the optimal information feedback in this regime of shallow gradients, it is useful to analyze the bounds that follow from our navigation-information equality, Eq. \ref{eq:smallkinfo}. In particular, Eq. \ref{eq:smallkinfo} implies that the performance is bounded by
\begin{align}
\frac{\mathcal{P}}{\mathcal{P}_{0}}\leq {\rm MIN} \{4 \mathcal{T}_{\mathrm{FF}} \sqrt{\mathcal{T}_{\mathrm{FB}}},\mathcal{T}_{\mathrm{FF}}, \frac{\mathcal{T}_{\mathrm{FF}}}{\mathcal{T}_{\mathrm{FB}}}\}. \label{eq:smallkbound}
\end{align}
These bounds are illustrated in Fig. \ref{fig:spatialinfo}A, which shows the performance as the feedback information is varied while the feedforward information is kept constant. In the regime of weak feedback, the performance is limited by the first argument of Eq. \ref{eq:smallkbound}: navigation depends on both the feedforward and feedback information, and a stronger feedback information improves navigation because the actuation system responds more accurately to the information provided by the sensory system. At the optimal feedback information $\mathcal{T}^{*}_{\mathrm{FB}}$, navigation is limited by the feedforward information alone, corresponding to the second argument of Eq. \ref{eq:smallkbound}; here, navigation can be improved only by improving the sensory system. For stronger feedback, navigation performance decreases with with increasing information feedback, as captured by the third argument of Eq. \ref{eq:smallkbound}. In this regime, the actuation system responds too sensitively to the sensory output, making navigation susceptible to sensory noise. Further mechanistic insight is obtained by analyzing the performance in terms of the elementary system parameters. This reveals that the optimal feedback information $\mathcal{T}^{*}_{\mathrm{FB}}$ is associated with an optimal actuation strength $J^{*}=F\sqrt{D_{v}/D_{f}}$: a higher sensory noise $D_f$ calls for weaker actuation. In the regime of shallow gradients, the optimal feedback information thus arises from the trade-off between responding to the sensory signal but not to sensory noise; the optimum is not associated with the phenomenon of concentration peak overshoots, seen for steeper gradients.

Eq. \ref{eq:fullP} shows that in shallow gradients, navigation is controlled by bidirectional information flow alone. But do additional constraints emerge for steeper gradients? For any value of gradient steepness $k$, we can express the navigation performance as
\begin{align}
\frac{\mathcal{P}}{\mathcal{P}_{0}}=\frac{\widetilde{\mathcal{T}}_{\mathrm{FF}}\widetilde{\mathcal{T}}_{\mathrm{FB}}\rho[1+\rho-\widetilde{\mathcal{T}}_{\mathrm{FF}}\widetilde{\mathcal{T}}_{\mathrm{FB}}]}{\rho(1+\rho)(1+\widetilde{\mathcal{T}}_{\mathrm{FB}}^{2})+\widetilde{\mathcal{T}}_{\mathrm{FF}}\widetilde{\mathcal{T}}_{\mathrm{FB}}},  \label{eq:allgradients}
\end{align}
where $\rho=F/H$, while 
$\widetilde{\mathcal{T}}_{\mathrm{FF}}=[(2\mathcal{T}_{\mathrm{FF}}+1)^{2}-1]/2$ and $\widetilde{\mathcal{T}}_{\mathrm{FB}}=\sqrt{(2\mathcal{T}_{\mathrm{FB}}+1)^{2}-1}$ are monotonically increasing functions of $\mathcal{T}_{\mathrm{FF}}$ and $\mathcal{T}_{\mathrm{FB}}$, respectively. In shallow gradients, where $\widetilde{\mathcal{T}}_{\mathrm{FF}}$ is small compared to $\rho$,
this expression indeed reduces to Eq. \ref{eq:smallkinfo}. For steeper gradients, however, this is no longer true: in this regime, navigation is controlled not only by the feedforward and feedback transfer entropies,
but also by the relative rate of information processing over actuation, $\rho$. 
 
In steep gradients, the fundamental collective variables that control navigation are thus $\mathcal{T}_{\rm{FF}}$, $\mathcal{T}_{\rm{FB}}$ and $\rho$. The implications are twofold. First, 
at fixed 
$\mathcal{T}_{\rm{FF}}$ and $\mathcal{T}_{\rm{FB}}$, the performance can only be raised by increasing  $\rho$, as shown in Fig. \ref{fig:spatialinfo}B. Secondly, 
the performance in steep gradients is bounded by
\begin{align}
\frac{\mathcal{P}}{\mathcal{P}_{0}}\leq\frac{\widetilde{\mathcal{T}}_{\mathrm{FF}}\widetilde{\mathcal{T}}_{\mathrm{FB}}\rho[1+\rho]}{\rho(1+\rho)(1+\widetilde{\mathcal{T}}_{\mathrm{FB}}^{2})}=\frac{\widetilde{\mathcal{T}}_{\mathrm{FF}}}{\left( \widetilde{\mathcal{T}}_{\mathrm{FB}}+\frac{1}{\widetilde{\mathcal{T}}_{\mathrm{FB}}}\right)},
\end{align}
which is the performance in shallow gradients for the same amount of bidirectional information. Therefore, the same amounts of feedforward and feedback information are less useful to the navigating cell in steeper gradients than in shallow gradients, unless the information is acquired and processed sufficiently rapidly.

\subsection*{Single-step transfer entropy does not fully determine performance} The utility of the bidirectional transfer entropy rates as fundamental variables for navigation crucially depends on accounting for the full history of the signal trajectories. If the history is instead truncated to only the current signal state, the resultant transfer entropy is called the single-step transfer entropy \cite{chetrite2019information}. 
We test the applicability of this truncation in the spatial-sensing model by analytically solving for the single-step transfer entropy rates. We find that the difference between the single-step and the full history-dependent transfer entropy rates can be large, and even diverge in the regime of weak actuation. 
Moreover, as we describe in the SI, we find that the navigation performance cannot be expressed in terms of these one-step transfer entropy rates alone.  
This demonstrates that the full signal history must be accounted for to understand how information controls navigation.

\subsection*{Minimal model for temporal sensing}
We next consider navigators that use temporal sensing. This strategy is intrinsically nonlinear: because the sensory output is a scalar and therefore lacks directional information, it cannot directly steer the cell’s velocity as in the spatial-sensing model. Instead, it modulates the rate at which the cell reorients, thereby biasing future motion only through the statistics of directional changes.
We therefore analyze a weakly nonlinear model that is inspired by the chemotaxis system of \textit{E. coli} \cite{long2017feedback}. To keep the model analytically tractable, we consider a constant chemoattractant gradient.

The minimal model consists of the sensory output $f$ and the velocity $v$ of a cell moving in a linear chemoattractant concentration profile $c(x)=gx$ (see Fig. \ref{fig:temporalmodel}A). The sensory input is the temporal derivative of the concentration along the cell's trajectory, $s(t)=dc/dt=(dc/dx).(dx/dt)=gv(t)$. The cell amplifies this sensory input linearly with sensory gain $G$, while actuation is described by a bilinear coupling between the sensory output and the velocity, with actuator gain $J$. This bilinear form for the coupling can be derived from a strongly nonlinear model of the chemotaxis system of \textit{E. coli} in the regime of small noise \cite{long2017feedback}. As in the spatial sensing-model, the sensory and actuator relaxation rates are denoted by $F$ and $H$, respectively, and the corresponding noise magnitudes by $D_{f}$ and $D_{v}$.

Despite the simplicity of the minimal model, it exhibits chemotaxis through nonlinear positive feedback, as observed in earlier models of \textit{E. coli} chemotaxis \cite{long2017feedback,flores2012signaling}. When the velocity is close to zero, the sensory input is weak and the sensory output stays close to zero. A positive fluctuation
in the velocity, however, generates a positive sensory input, causing the sensory output $f$ to increase. Through the bilinear coupling between $f$ and $v$, this positive sensory output provides an additional positive drive on the velocity. Positive velocity fluctuations are therefore amplified and persist longer than negative fluctuations. This asymmetry generates a net drift up the gradient, which we take as the navigation performance, $\mathcal{P}=\langle v\rangle$. Typical trajectories are shown in Fig. \ref{fig:temporalmodel}B, where $v(t)$ displays large positive excursions generated by this positive feedback loop. \

To compare the drift speed to the bidirectional transfer entropy, we again need to account for their different physical dimensions, as in the spatial-sensing model. We express the navigation performance $\mathcal{P}$ in units of the typical velocity fluctuations of the system in the absence of actuation, $\mathcal{P}_{0}=\sigma_{v0}=\sqrt{D_{v}/H}$. We define the feedforward and feedback information by multiplying  the corresponding transfer entropy rates  by the memory timescales of their respective input signals. With this construction, we begin deriving a Behavioral Equation of State for the temporal-sensing model.

\subsection*{Nonlinear perturbation theory for information and performance}\label{sec:perturb}
To derive analytical expression for information transmission and navigation performance in the temporal-sensing model, we develop a controlled perturbation theory for the regime of weak nonlinearity. 
Such an approach is necessary because  transfer entropy can be computed analytically in closed form for linear stochastic systems, but generally not for nonlinear systems \cite{tostevin2009mutual,chetrite2019information}. Recent approximate approaches for nonlinear systems are promising, but rely on heuristic approximations that can lead to uncontrolled errors \cite{moor2023dynamic}. To obtain controlled analytical expressions up to a prescribed order in the nonlinearity, we combine perturbation theory from statistical physics with stochastic filtering theory.

Stochastic filtering theory is directly applicable to the computation of  transfer entropy rates \cite{chetrite2019information, moor2023dynamic}. The theory describes how an input signal can be reconstructed from the trajectory of a noisy observation of that input \cite{aastrom2012introduction}. It therefore provides a natural framework for quantifying the information contained in the sensory output about the sensory input, and conversely the information contained in the sensory input about the sensory output through the actuation loop.
In linear systems, filtering theory yields a Riccati equation that can be solved to obtain the transfer entropy rate. We extend this approach to systems with bilinear coupling such as our minimal model for temporal sensing. This leads to a stochastic Riccati equation, which we solve perturbatively to derive analytical expressions for the transfer entropy rates and navigation performance up to second order in the nonlinear coupling, the actuator gain $J$ (see Materials and Methods).

\begin{figure*}[t!]
\centering
\includegraphics[trim={0cm 0 0 0},clip,width=11.4cm]{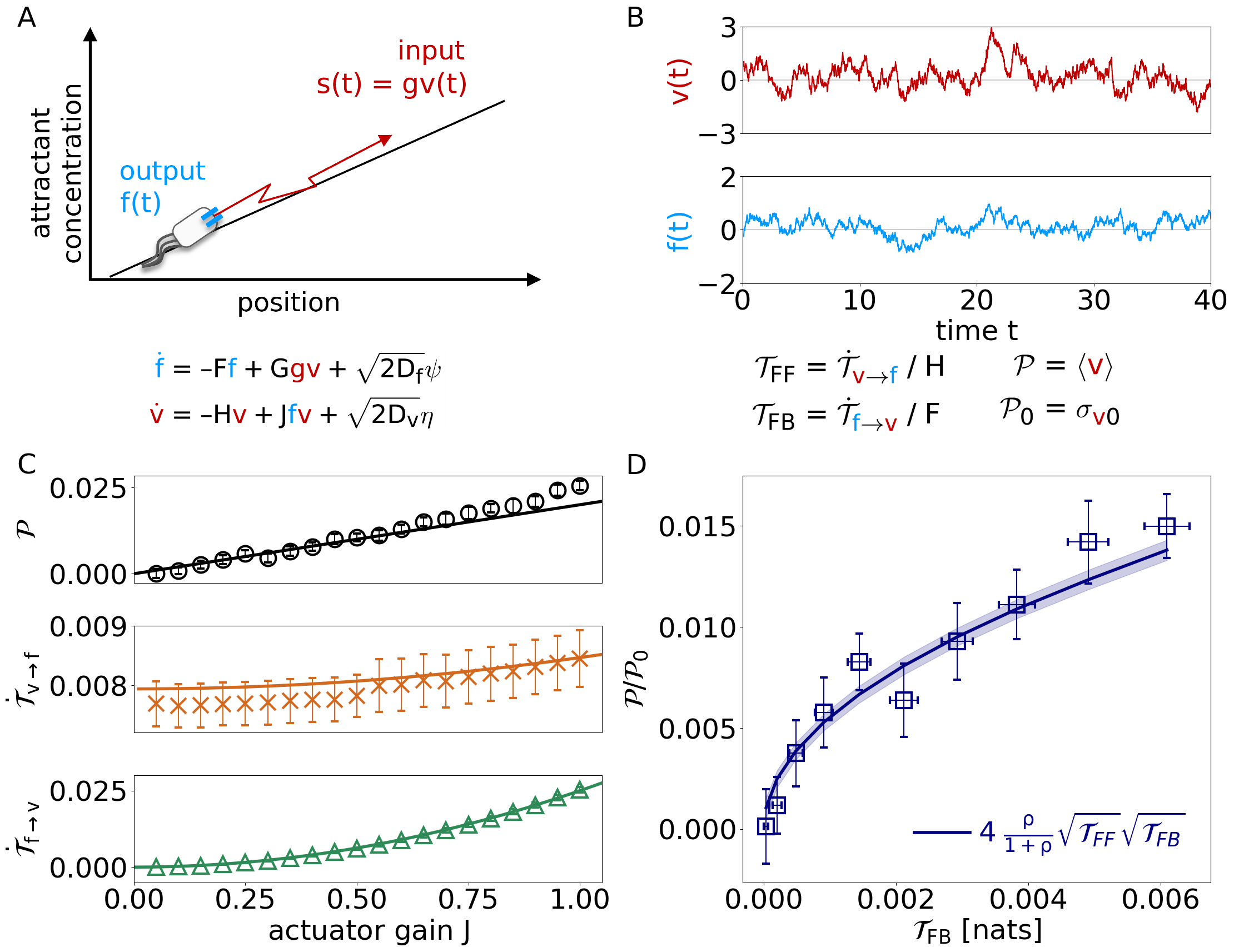}
\caption{In the temporal-sensing model, navigation performance is controlled by bidirectional information transmission.
(A) Schematic illustration of temporal-sensing model. The sensory input is the product of the gradient strength, $g$, and the cell velocity, $v$. The sensory system maps the input onto the sensory output $f$, with sensory gain $G$ and sensory noise $D_{f}$. The sensory output then drives the actuator to modulate the cellular velocity, with actuator gain $J$ and actuation noise $D_{v}$. The variables $f$ and $v$ relax to zero with rates $F$ and $H$, respectively. (B) Representative trajectories show gradient climbing through positive feedback. Upper panels show a representative trajectory of the model for $F=J=H=1,Gg=0.08,D_{f}=0.1,D_{v}=0.5$.
The trajectories show large excursions, due to the positive feedback between $f$ and $v$. The lower panel defines the normalized performance and information metrics. (C) Nonlinear perturbation theory accurately predicts information transmission and performance. Solid lines are the analytical expressions; symbols correspond to simulation results obtained from the minimal model using the TE-PWS algorithm. (D) Temporal-sensing cells obey the Behavioral Equation of State (BEST). The solid line is the theoretical prediction for the performance from the measured feedforward and feedback information, $\mathcal{P}/\mathcal{P}_0 = 4[\rho/(1+\rho)]\sqrt{\mathcal{T}_{\rm{FF}}}\sqrt{\mathcal{T}_{\rm{FB}}}$, where $\mathcal{T}_{\rm{FF}}$ and $\mathcal{T}_{\rm{FB}}$ are obtained from the numerical simulations using TE-PWS. The uncertainty in the solid line originates from statistical errors in the TE-PWS estimates of $\mathcal{T}_{\rm{FF}}$ and $\mathcal{T}_{\rm{FB}}$. Symbols show the navigation performance measured directly from the numerical simulations.}
\label{fig:temporalmodel}
\end{figure*}

We test the validity of the analytical expressions by comparing them to numerically exact values obtained from simulations of the same minimal model, see Fig. \ref{fig:temporalmodel}C. The exact transfer entropy rates are computed using the TE-PWS algorithm \cite{das2025exact}. In the regime of weak nonlinearity, corresponding to small values of $J$, both the navigation performance and the information measures are accurately described by the analytical expressions. As the nonlinearity increases, the performance begins to deviate from the perturbative theory. In the limit of large $J$, the minimal model becomes unstable because of runaway positive feedback between $f$ and $v$. Biological systems avoid such instabilities through resource-imposed bounds on the maximum attainable values of $f$ and $v$, similar to the spatial-sensing model.

Before we analyze whether navigation performance is controlled by bidirectional information flow alone, we consider the analytical expressions for the transfer entropy rates and the navigation performance in terms of the elementary system parameters. These demonstrate several non-trivial consequences of the nonlinear coupling. The feedforward transfer entropy rate is $\dot{\mathcal{T}}_{v\to f}=[\sqrt{H^{2}+G^{2}g^{2}D_{v}/D_{f}}-H]/2+J^{2}\Delta \mathcal{T} +\mathcal{O}(J^{3})$, with the expression for $\Delta \mathcal{T}$ given in the Materials and Methods section. Importantly, unlike in the spatial-sensing model, the feedforward information depends on the feedback coupling strength $J$ (see Fig. \ref{fig:temporalmodel}C).  The reason is that the positive feedback  between $f$ and $v$ enables the actuation to change the statistics of the sensory input, thereby improving sensory information transmission.

In contrast, the feedback transfer entropy rate, $\dot{\mathcal{T}}_{f\to v}=J^{2}D_{f}/(4HF)+\mathcal{O}(J^{3})$, is independent of the feedforward gain $G$ to this order. Moreover, unlike in the spatial-sensing model, this transfer entropy rate is independent of the actuation noise magnitude $D_{v}$. This is because actuation noise has two opposing effects. On the one hand, the precision of actuation is adversely affected by actuation noise. On the other hand, this very noise induces the positive feedback loop that generates large excursions in $v$, thereby increasing feedback information. These two effects cancel at this order, making the feedback information insensitive to actuation noise.

Lastly, the drift speed is linear in the actuator gain $J$, $\mathcal{P}=\langle v\rangle = JGgD_{v}/[H^{2}(H+F)]+\mathcal{O}(J^{3})$. Therefore, owing to the positive feedback loop, the drift speed is amplified by the actuator noise $D_{v}$, again in marked contrast to spatial sensing.

Despite these differences between spatial and temporal sensing, the central challenge remains the same: navigation performance and information transmission depend in a highly nontrivial way on the elementary system parameters. We next show that, as in the spatial-sensing model, navigation can be expressed in terms of information measures alone.

\subsection*{Performance depends solely on bidirectional information transmission in temporal sensing}
In the regime of shallow gradients and weak actuation, where our perturbation theory applies, we analytically find that the performance is fully determined by the amount and speed of bidirectional information transmission,
\begin{align}\label{eq:temporalbest}
\frac{\mathcal{P}}{\mathcal{P}_{0}}&=4\frac{\rho}{1+\rho}\sqrt{\mathcal{T}_{\mathrm{FF}}}\sqrt{\mathcal{T}_{\mathrm{FB}}}.
\end{align}
This is our Behavioral Equation of State for temporal sensing. It shows that, as for spatial sensing, navigation is controlled by bidirectional information transmission alone
in the regime of weak actuation and shallow gradients. The elementary system parameters affect navigation only through their effects on information transmission.
Even more strikingly, the expression is similar to that obtained for spatial sensing in this regime, $\mathcal{P}/\mathcal{P}_{0}=4\mathcal{T}_{\mathrm{FF}}\sqrt{\mathcal{T}_{\mathrm{FB}}}$ (Eq. \ref{eq:smallkbound}). In both cases,  the performance scales as the square root of the feedback information.

There are also two important differences. First, for temporal sensing navigation performance scales as the square root of the feedforward information, rather than linearly as in spatial sensing (Eqs. \ref{eq:smallkinfo} and \ref{eq:smallkbound}). This is a fundamental consequence of the temporal sensing strategy: the sensory input depends not only on the gradient strength but also on the cellular motion, while the latter itself also depends on the gradient strength \cite{mattingly2021escherichia}. Second, the relative speed of sensory processing compared to feedback, $\rho=F/H$, affects the performance even in the regime of shallow gradients; performance increases monotonically with $\rho$. This is because the sensory input, given by the velocity $v$, fluctuates on the timescale of $1/H$, regardless of the fact that the gradient is shallow. The sensory system must therefore transmit information faster than this timescale to avoid errors due to dynamical lag \cite{tjalma2023trade}. In Fig. \ref{fig:temporalmodel}D, we numerically verify that the numerically exact results from simulations of the minimal model obey the BEST in the regime of shallow gradients and weak actuator gain.

\subsection*{BEST predicts chemotaxis in \textit{E. coli}}
We test the validity of BEST by performing stochastic simulations of a realistic model of the chemotaxis system of \textit{E. coli}. {\it E. coli} performs chemotaxis by alternating between runs and tumbles, and uses a temporal-sensing strategy to modulate the switching propensities between these two motility states. When the bacterium swims up the gradient, the temporal derivative of the concentration is positive. This changes the sensory output and decreases the propensity to switch from running to tumbling, thereby prolonging runs in the favorable direction. Conversely, when the bacterium swims down the gradient, the switching propensity increases. We therefore define the dependence of the run-to-tumble switching propensity on the sensory output as the actuator gain.
In the simulations, we vary this gain to modulate information transmission through the actuator, allowing us to test how navigation performance depends on actuator-driven feedback information.

We perform agent-based simulations of  \textit{E. coli} using a coarse-grained model with three variables: the sensory output $f(t)$, the projection of the swimming direction on the gradient direction, $n(t)$, and the motility state $\delta_{R(t)}$, which equals 1 when the cell is running and 0 when it is tumbling.  
This model is adopted from a previously developed model \cite{long2017feedback}. The cell is placed in an exponential concentration profile $c(t)=c_{0}e^{gx(t)}$, which corresponds to a linear profile with slope $g$ to log-sensing organisms such as \textit{E. coli} \cite{shimizu2010modular}: the sensory input is the temporal derivative of the logarithmic concentration, given by $d\log c / dt = gv(t)=gv_{0}n(t)\delta_{R(t)}$, where $v_{0}$ is the run speed of the cell. This input drives sensing and actuation through nonlinear equations of motion,
\begin{align}
\dot{f}&=-F(f-f_{0})+Ggv_{0}n\delta_{R}+\sqrt{2D_{f}}\psi,\label{eq:Ecolif}\\
\lambda_{\mathrm{R}\to\mathrm{T}}(t)&=\omega_{_{R}}\mathrm{max}\{ [1-J(f-f_{0})],0\},\\
\lambda_{\mathrm{T}\to\mathrm{R}}(t)&=\omega_{T},\\
\dot{n}&=-n/(2D(\delta_R))+\sqrt{(1-n^{2})/(2D(\delta_R))}\;\eta.\label{eq:Ecolin}
\end{align}
Here, $F$ is the adaptation rate, $G$ the sensory gain, $J$ the actuator gain, $D_{f}$ the sensory noise,
$\lambda_{\mathrm{R}\to\mathrm{T}}$ and $\lambda_{\mathrm{T}\to\mathrm{R}}$ are the switching propensities from run to tumble and from tumble to run, respectively,  $D(\delta_R)$ is the orientational diffusion constant of the cell, which depends on the motility state of the cell, $\delta_R$, and $\psi$ and $\eta$ are Gaussian white noise sources.
Values of all model parameters are adopted from \cite{long2017feedback} and are tabulated in the Materials and Methods section. 

In our theory, the relation between navigation and information depends on a single timescale ratio, $\rho=F/H$ (Eq. \ref{eq:temporalbest}). Here, $F$ is the sensory information processing speed, which appears in Eq. \ref{eq:Ecolif}, and $H$ is the rate at which velocity fluctuations decay. In contrast, the agent-based model of Eqs. \ref{eq:Ecolif}-\ref{eq:Ecolin} contains two distinct motility timescales, associated with motor switching and orientational diffusion. This could make the definition of $H$ ambiguous. To avoid this ambiguity, we exploit the phenotypic diversity of {\it E. coli}, which gives rise to a distribution of wild-type values for the orientational diffusion constants \cite{long2017feedback}. From this distribution, we select values close to the median for which the velocity autocorrelation function is well described by a single exponential decay, thereby defining a single velocity correlation timescale $H^{-1}$.

\begin{figure}[t]
\centering
\includegraphics[trim={0 0 0 0cm},clip,width=8.7cm]{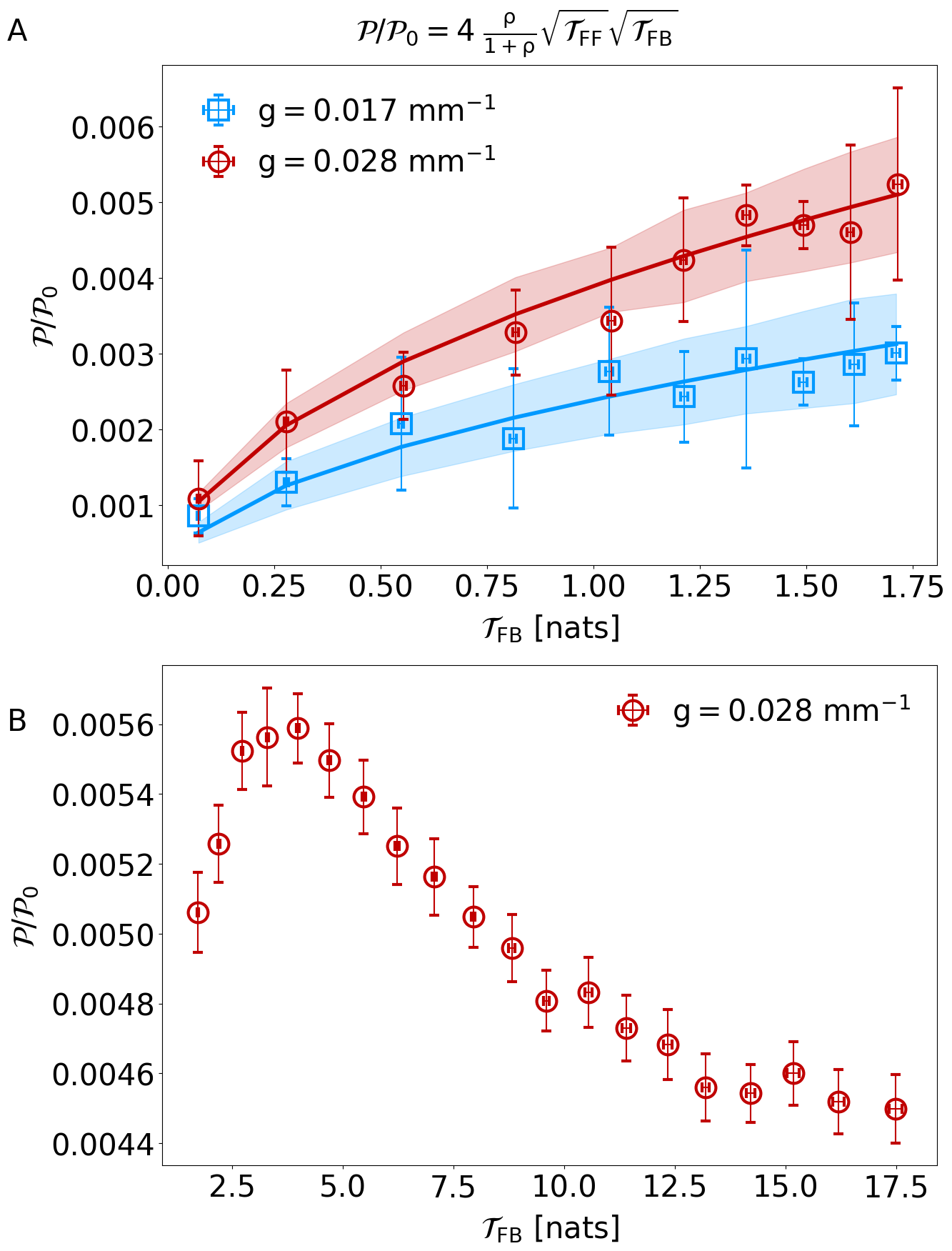}
\caption{BEST predicts \textit{E. coli} chemotaxis. (A) Navigation performance as a function of feedback information, for two different gradient strengths $g$. In each case, the solid line is the BEST prediction for the performance from the measured feedforward and feedback information in the regime of shallow gradients and weak actuation, $\mathcal{P}/\mathcal{P}_0 = 4[\rho/(1+\rho)]\sqrt{\mathcal{T}_{\rm{FF}}}\sqrt{\mathcal{T}_{\rm{FB}}}$, where $\mathcal{T}_{\rm{FF}}$ and $\mathcal{T}_{\rm{FB}}$ are obtained from the simulations using TE-PWS while varying 
the actuator gain $J$. The uncertainty in the solid line originates from statistical errors in the TE-PWS estimates for $\mathcal{T}_{\rm{FF}}$ and $\mathcal{T}_{\rm{FB}}$. 
Because $\mathcal{T}_{\rm FF}$ is small and independent of $J$ within the regime considered, the values of  $\mathcal{T}_{\rm FF}$ for different $J$ were averaged to improve the estimate. 
Symbols show the navigation performance measured directly from the numerical simulations. (B) For larger feedback information, the performance exhibits a turnover, revealing an optimal feedback information that maximizes navigation performance.}
\label{fig:ecoli}
\end{figure}

To test the validity of the temporal-sensing BEST, Eq. \ref{eq:temporalbest}, in this realistic model, we measure $\mathcal{P}$, $\mathcal{T}_{\mathrm{FF}}$, $\mathcal{T}_{\mathrm{FB}}$ and $\rho$ directly from agent-based simulations. Because the sensory input is jointly determined by the run/tumble state of the cell and the swimming direction, we compute the bidirectional transfer entropy rates between the joint trajectory of $(n,R)$ and the trajectory of the sensory output $f$, using the TE-PWS algorithm \cite{das2025exact}. We measure $H$ as the decay rate of the velocity autocorrelation function in simulations with $J=0$, while
$F$ is directly specified by the model. This gives the dimensionless information measures $\mathcal{T}_{\mathrm{FF}}=\dot{\mathcal{T}}_{(n,R)\to f}/H$ and $\mathcal{T}_{\mathrm{FB}}=\dot{\mathcal{T}}_{f\to(n,R)}/F$, together with $\rho=F/H$. We then compare the relationship between these quantities and navigation performance with the prediction of Eq. \ref{eq:temporalbest}.

We find that our analytical theory predicts the chemotactic performance of \textit{E. coli} without any fitting or scaling parameters, in the regime of shallow gradients and weak actuation (see Fig. \ref{fig:ecoli}A). The amount and speed of the bidirectional information transmission fully determine navigation performance in this regime. As we vary the actuator gain $J$ to modulate the feedback information $\mathcal{T}_{\mathrm{FB}}$, the performance changes as predicted by Eq. \ref{eq:temporalbest}. Likewise, varying the gradient strength changes the  feedforward information $\mathcal{T}_{\mathrm{FF}}$, and thereby changes the performance as predicted by Eq. \ref{eq:temporalbest}.
This strongly suggests that the Behavioral Equation of State, Eq. \ref{eq:temporalbest}, captures a general principle for temporal-sensing stochastic navigators in the regime of shallow gradients and weak actuation, independent of the microscopic details of their sensory and actuation systems.

Finally, inspired by the optimum in the feedback information in the spatial-sensing model, we numerically tested whether such an optimum arises in \textit{E. coli} chemotaxis (see Fig. \ref{fig:ecoli}B). Our perturbative theory is only applicable to the regime of weak actuation, and therefore cannot analytically predict either the existence or the position of such an optimum. However, the agent-based simulations show that as the feedback information is varied, the performance exhibits a turnover, with an optimum of around $4\mathrm{~nats}$. This suggests that the qualitative design principle identified in the spatial-sensing model extends to realistic temporal-sensing navigators: excessive feedback information can become detrimental to function, because an ultrasensitive actuator amplifies sensory noise.

\section*{Discussion}
We have shown that, for two distinct navigation strategies, spatial and temporal sensing, navigation performance is controlled, at least in the linear-response regime of shallow gradients, by bidirectional information transmission alone. System-specific parameters, such as sensory gain, sensory noise, and actuation strength, affect navigation only through their effects on the amount and speed of information transmission. This suggests that information provides a universal organizing principle for stochastic navigation, at least within this regime. Importantly, the same relation also predicts the chemotactic performance of a realistic, nonlinear model of {\it E. coli} without any fitting or scaling parameters.

Recently, Chen et al. independently studied how navigation is shaped by bidirectional information flow \cite{chen2026behavior}. They developed a coarse-grained four-state model suited for experimental data analysis, derived a scaling relation between navigation performance and bidirectional information flow, and tested it on both synthetic and experimental time-series data. Like us, they conclude that bidirectional information drives navigation. However, the two approaches differ in key respects. First, whereas our theory uses history-dependent transfer entropy, their framework relies on single-step transfer entropy. 
Second, both the feedforward information and, more importantly, the feedback information are defined differently. Navigation requires that changes in the sensory output are transmitted through the actuation system to alter the future sensory input. Our definition of feedback information quantifies this closed-loop causal link, from sensory output through actuation back to future input. With this definition, navigation requires nonzero feedback information, whereas in the framework of Chen et al. navigation can occur with zero feedback information.

The equalities we have derived between information transmission and navigation performance predict an
experimentally testable data collapse. Specifically, data from different cells or mutants, with different sensing and actuation parameters, such as sensory gain, sensory noise, or actuation strength, 
should collapse onto the same BEST curve, which specifies the relation between navigation performance and bidirectional information. The required bidirectional information flows can be obtained from experimental time-series data using  the recently developed ML-PWS algorithm \cite{reinhardt2025ml}.

Although we studied the spatial and temporal sensing strategies for two different navigation objectives, the corresponding BEST relations are strikingly similar. In the regime of shallow gradients, weak actuation and fast sensing ($\rho\gg 1$), the normalized performance is $4\mathcal{T}_{\mathrm{FF}}\sqrt{\mathcal{T}_{\mathrm{FB}}}$ for spatial sensing, and $4\sqrt{\mathcal{T}_{\mathrm{FF}}}\sqrt{\mathcal{T}_{\mathrm{FB}}}$ for temporal sensing. Thus, the only difference between the two is the dependence on the feedforward information $\mathcal{T}_{\mathrm{FF}}$. This difference, however, stems from the fundamental difference in how the gradient is estimated, not the navigation objective. It raises the question of whether the BEST relation also holds for other navigation objectives. We leave this for future work.

In our work, we have analyzed idealized environments, such as harmonic concentration peaks and (effectively) linear concentration gradients. However, real environments are likely to be much more complex. Living cells often navigate complex environments, such as turbulent flows, porous soils, or concentration profiles shaped by other cells \cite{taylor2012trade,de2021chemotaxis,saragosti2011directional}. Such environments generate sensory input signals with multiple timescales. Our spatial-sensing model already shows that, while navigation in shallow gradients is controlled by the amount of bidirectional information, steeper gradients also require information to be acquired and processed sufficiently rapidly. This suggests that, in more complex environments, additional collective variables will become important, such as ratios of timescales in sensing, actuation, and environmental variation. A key question is whether navigation performance in such environments can still be predicted from a limited set of collective variables: bidirectional information transmission, together with additional variables that capture the temporal structure of sensing, actuation, and environmental variation.

Finally, stochastic feedback between sensing and actuation is central to many biological systems, ranging  from cellular homeostasis \cite{Aoki.2019}, immune response \cite{Ukogu.2026}, and virus spreading \cite{meijers2023population}, to social interactions \cite{o2024dynamics}. In all these cases, the response reshapes the future input experienced by the system, creating closed loops of bidirectional information flow  between input sensing and actuation. It is tempting to speculate that other feedback-driven biological systems also obey relations between function and information analogous to the Behavioral Equation of State identified here.

\section*{Materials and Methods}
\subsection*{Derivation of information and performance measures in spatial-sensing model}
We here outline the derivation of the analytical expressions for the information and the performance in the minimal model for spatial sensing. A full derivation is provided in the SI Appendix. Simplifications of the analytical expressions were performed with Wolfram Mathematica 14.0.

The minimal model for spatial sensing is reproduced for convenience,
\begin{align}
\dot{x}&=v\label{eq:spatialx},\\
\dot{f}&=-Ff+G.(-kx)+\sqrt{2D_{f}}\psi\label{eq:spatialf},\\
\dot{v}&=-Hv+Jf+\sqrt{2D_{v}}\eta\label{eq:spatialv}.
\end{align}
We first note that since $\dot{x}$ is deterministic in $v$, the trajectories of $x$ and $v$ can be mapped onto each other one-to-one. Therefore, the full trajectory-dependent transfer entropy rates between $f$ and $x$ are the same as between $f$ and $v$. We solve for the latter for analytical convenience.

For solving for $\dot{\mathcal{T}}_{f\to v}$ and $\dot{\mathcal{T}}_{v\to f}$, we extend an existing analytical approach based on stochastic control theory \cite{chetrite2019information} from two to three dimensions. Below we describe the approach for the feedback transfer entropy rate $\dot{\mathcal{T}}_{f\to v}$. The derivation for the feedforward transfer entropy rate $\dot{\mathcal{T}}_{v\to f}$ is analogous.

The trajectories of Eqs. \ref{eq:spatialx}-\ref{eq:spatialv} are Gaussian distributed since the dynamics is linear. Therefore, the transfer entropy rate is
\begin{align}\label{eq:transfervar}
\dot{\mathcal{T}}_{f\to v}=\lim_{\delta t\to 0+}\frac{1}{2\delta t} \log\frac{\sigma^{2}_{v(t+\delta t)|\mathbf{v}_{[-\infty,t]}}}{\sigma^{2}_{v(t+\delta t)|\mathbf{f}_{[-\infty,t]},\mathbf{v}_{[-\infty,t]}}},
\end{align}
where $\sigma^{2}_{v(t+\delta t)|\mathbf{v}_{[-\infty,t]}}$ and $\sigma^{2}_{v(t+\delta t)|\mathbf{f}_{[-\infty,t]},\mathbf{v}_{[-\infty,t]}}$ are conditional variances for $v(t+\delta t)$ within two marginal spaces: the space of only $v$, where both $x$ and $f$ have been marginalized out, and the space of $(f,v)$, where only $x$ has been marginalized out, respectively

For obtaining the two variances, we need to solve for the trajectory of $v(t)$ within these two marginal spaces. The \textit{effective dynamics} within the marginal spaces are defined through the respective \textit{causal functions} $H^{(v)}(t)$ and $\mathbf{H}^{(fv)}(t)$.
$H^{(v)}(t)$ is the causal function that generates the exact distribution of the $v$ trajectory from a Gaussian white noise $\xi^{(v)}(t)$ with variance $2D_{v}$, through the equation
\begin{align}\label{eq:veq1}
v(t)=\int_{-\infty}^{t}ds~H^{(v)}(t-s)\xi^{(v)}(s).
\end{align}
Similarly, $\mathbf{H}^{(fv)}(t)$ is the causal matrix that generates the exact joint distribution of the $f$ and $v$ trajectories from Gaussian white noise $\zeta^{(f)}(t)$ and $\zeta^{(v)}(t)$ with variances $2D_{f}$ and $2D_{v}$, through the equations
\begin{align}
f(t)&=\int_{-\infty}^{t}ds~\left[ H^{(fv)}_{ff}(t-s)\zeta^{(f)}(s)+H^{(fv)}_{fv}(t-s)\zeta^{(v)}(s)\right] ,\\
v(t)&=\int_{-\infty}^{t}ds~\left[ H^{(fv)}_{vf}(t-s)\zeta^{(f)}(s)+H^{(fv)}_{vv}(t-s)\zeta^{(v)}(s)\right]  .\label{eq:veq2}
\end{align}

Solving for the transfer entropy rate then reduces to solving for the causal functions. This is achieved in Fourier space through a Wiener-Hopf factorization of the power spectrum of the system. For example, $H^{(v)}(\omega)$, the Fourier transform of $H^{(v)}(t)$, can be solved from the equation
\begin{align}
S_{vv}(\omega)=H^{(v)}(\omega).2D_{v}.H^{(v)}(-\omega),
\end{align}
where $S_{vv}(\omega)$ is the power spectrum of $v(t)$. We achieve this factorization analytically. This gives us the causal functions. 

The causal functions can then be used to obtain the variances of Eqs. \ref{eq:veq1} and \ref{eq:veq2}. This, in turn, produces the transfer entropy rate.

The navigation performance, $\mathcal{P}=1/\sigma_{x}^{2}$, is obtained from the stationary covariance matrix $\mathbf{C}$ of the model. 
$\mathbf{C}$ is derived by analytically solving the Lyapunov equation $\mathbf{AC}+\mathbf{CA}^{T}=2\mathbf{D}$, where $\mathbf{A}=\begin{pmatrix}0&0&-1\\Gk&F&0\\0&-J&H\end{pmatrix}$ and $\mathbf{D}=\begin{pmatrix}0&0&0\\0&D_{f}&0\\0&0&D_{v}\end{pmatrix}$.

\subsection*{Nonlinear perturbation theory for information and performance measures in temporal-sensing model}
We here outline the derivation of the perturbation theory in the minimal model for temporal sensing. A full derivation is provided in the SI Appendix. The conceptual framework of the theory was proposed by the authors. The analytical expressions were derived through iterative interactions with Claude Sonnet 4.6 and ChatGPT-5.5 Pro, and verified independently by the authors.

We reproduce the temporal sensing model here for convenience,
\begin{align}
\dot{f}&=-Ff+Ggv+\sqrt{2D_{f}}\psi\label{eq:tempmodel1},\\
\dot{v}&=-Hv+Jfv+\sqrt{2D_{v}}\eta\label{eq:tempmodel2}.
\end{align}
Since this is a diffusion process, we can apply the Girsanov formula for path measures to express the transfer entropy rates as \cite{girsanov1960transforming,chetrite2019information}
\begin{align}
\dot{\mathcal{T}}_{v\to f}&=\frac{G^{2}g^{2}}{4D_{f}}\left\langle P^{v|f}\right\rangle\label{eq:girsanov1},\\
\dot{\mathcal{T}}_{f\to v}&=\frac{J^{2}}{4D_{v}}\left\langle v^{2}P^{f|v}\right\rangle\label{eq:girsanov2},
\end{align}
where $P^{f|v}=\mathrm{Var}\left( P(f(t)|\mathbf{v}_{[-\infty,t]})\right)$ and $P^{v|f}=\mathrm{Var}\left( P(v(t)|\mathbf{f}_{[-\infty,t]})\right)$ are conditional variances. The angular brackets $\langle\;\cdot\;\rangle$ denote a steady-state average over the conditioned variable, in each case.

Since the nonlinear coupling term is, in fact, bilinear, the conditional probability distributions $P(f(t)|\mathbf{v}_{[-\infty,t]})$ and $P(v(t)|\mathbf{f}_{[-\infty,t]})$ are still Gaussian. They satisfy the \textit{stochastic} Riccati equations \cite{jazwinski2007stochastic},
\begin{align}
\dot{P}^{v|f}&=2[-H+Jf(t)]P^{v|f}+2D_{v}-\frac{G^{2}g^{2}}{2D_{f}}(P^{v|f})^{2}\label{eq:sr1},\\
\dot{P}^{f|v}&=-2FP^{f|v}+2D_{f}-\frac{J^{2}v(t)^{2}}{2D_{v}}(P^{f|v})^{2}\label{eq:sr2}.
\end{align}

In the limit of the actuator gain $J=0$, the above equations can be exactly solved. For solving them at a small non-zero value of $J$, we construct a perturbation expansion in $J$ for all the stochastic variables in the above equations:
\begin{align}
P^{v|f}(t)&=P_{0}^{v|f}(t)+JP_{1}^{v|f}(t)+J^{2}P_{2}^{v|f}(t)^{2}+\dots\\
P^{f|v}(t)&=P_{0}^{f|v}(t)+JP_{1}^{f|v}(t)+J^{2}P_{2}^{f|v}(t)^{2}+\dots\\
f(t)&=f_{0}(t)+Jf_{1}(t)+J^{2}f_{2}(t)+\dots\\
v(t)&=v_{0}(t)+Jv_{1}(t)+J^{2}v_{2}(t)+\dots\label{eq:vperturb}.
\end{align}
To solve for the unknown terms at each order of perturbation, we substitute these expansions into the stochastic Riccati equations \ref{eq:sr1} and \ref{eq:sr2} as well as the Langevin equations \ref{eq:tempmodel1} and \ref{eq:tempmodel2}. This yields self-consistent equations that are exactly solvable (see SI Appendix). This allows us to obtain analytical expressions for $P^{v|f}(t)$ and $P^{f|v}(t)$ up to a given order in $J$. 

\begin{table}[t!]
\centering
\caption{Model parameters for \textit{E. coli} adopted from [39]}
\begin{tabular}{cc}\label{tab:parameters}
Parameter & Numerical Value \\
\midrule
$f_{0}$ & 1.39 \\
$F$ & $0.1~s^{-1}$ \\
$G$ & $29.4$\\
$v_{0}$ & $20~\mu m~s^{-1}$\\
$D_{f}$ & $0.096~s^{-1}$\\
$\omega_{R}$ & $0.55~s^{-1}$\\
$\omega_{T}$ & $3.05~s^{-1}$\\
$D_{\mathrm{run}}$ & $0.062~s^{-1}$\\
$D_{\mathrm{tumble}}$ & $37D_{\mathrm{run}}$\\
\bottomrule
\end{tabular}
\end{table}

Finally, substituting the resultant solutions for $P^{v|f}(t)$ and $P^{f|v}(t)$ into Eqs. \ref{eq:girsanov1} and \ref{eq:girsanov2} yields the analytical expressions for the transfer entropy rates.

The derivation of the performance $\langle v\rangle$ is similar to that of the transfer entropy rates. Substituting Eq. \ref{eq:vperturb} in Eq. \ref{eq:tempmodel2} and taking the average relates the first order correction in $\langle v\rangle$ to the unperturbed stationary covariance $\langle fv\rangle_{0}$; the second order correction vanishes due to symmetry. The average $\langle fv\rangle_{0}$ is an element of the stationary covariance matrix $\mathbf{C}_{0}$ at $J=0$. $\mathbf{C}_{0}$ is derived by analytically solving the Lyapunov equation $\mathbf{AC}_{0}+\mathbf{C}_{0}\mathbf{A}^{T}=2\mathbf{D}$, where $\mathbf{A}=\begin{pmatrix}F&-Gg\\0&H\end{pmatrix}$ and $\mathbf{D}=\begin{pmatrix}D_{f}&0\\0&D_{v}\end{pmatrix}$. 

\subsection*{Analytical expressions for transfer entropy rates in temporal-sensing model}
The quantity $\Delta\mathcal{T}$, used in the perturbative expression of $\dot{\mathcal{T}}_{v\to f}$ in the section `Nonlinear perturbation theory for information and performance', is given by
\begin{align}
\Delta\mathcal{T}&=\frac{D_{f}(S^{2}-H^{2})(H^{2}F-H^{2}S+2FS^{2}+4S^{3})}{2H^{2}FS(F+H)(F+2S)(H+2S)},
\end{align}
where $S$ is defined as $\sqrt{H^{2}+G^{2}g^{2}D_{v}/D_{f}}$.

\subsection*{Chemotaxis in \textit{E. coli}}
Our model for \textit{E. coli} chemotaxis is adopted from the model developed in \cite{long2017feedback}, which has parameter values estimated from the experiments in \cite{shimizu2010modular, sneddon2012stochastic,berg1972chemotaxis,saragosti2012modeling,taheri2015cell}. 
The variable $f$, the sensory output, mimics the activity of the signaling protein CheY, which depends on the free-energy difference between the active and inactive state of the cellular receptors, which in turn depends on their ligand-binding state and methylation level. The values of the elementary parameters of the model are given in Table \ref{tab:parameters}.

\textbf{Data availability.}
Data and code for reproducing all figures in the manuscript are available on Zenodo \cite{datalink}.

\textbf{Acknowledgment.} We thank Vahe Galstyan, Daan Mulder, and the Emonet group for useful discussions. This work is part of the Dutch Research Council (NWO) and was performed at the research institute AMOLF. This project has received funding from the European Research Council under the European Union’s Horizon 2020 research and innovation program (grant agreement No. 885065). A.D. acknowledges start-up funding from Durham University.

\appendix

\end{document}


\beginsupplement

\title{Supporting Information: \\Navigation driven by bidirectional information transmission between sensing and actuation}

\preprint{APS/123-QED}
\author{Avishek Das}
\email{avishek.das@durham.ac.uk}
\affiliation{Department of Physics, Durham University, South Road, Durham DH1 3LE, UK\looseness=-1}
\author{Pieter Rein ten Wolde}
\email{p.t.wolde@amolf.nl}
\affiliation{AMOLF, Science Park 104, 1098 XG, Amsterdam, The Netherlands\looseness=-1}

\date{\today}

\maketitle

\thispagestyle{empty}
\onecolumngrid



\setcounter{equation}{0}
\setcounter{figure}{0}
\setcounter{table}{0}
\makeatletter
\renewcommand{\theequation}{S\arabic{equation}}
\renewcommand{\thefigure}{S\arabic{figure}}

\section*{Spatial sensing: linear theory for information and performance}

In order to derive analytical expressions for the information and performance in the minimal model for spatial sensing, we extend an existing theoretical approach in \cite{chetrite2019information} from two to three dimensions. This approach uses the linearity of the dynamics of the model.

The minimal model for spatial sensing is,
\begin{align}
\dot{x}&=v\label{eqs:spatialx},\\
\dot{f}&=-Ff+G.(-kx)+\sqrt{2D_{f}}\psi\label{eqs:spatialf},\\
\dot{v}&=-Hv+Jf+\sqrt{2D_{v}}\eta\label{eqs:spatialv}.
\end{align}
Since the trajectories of $x$ and $v$ deterministically map onto each other, the bidirectional transfer entropy rates between $x$ and $f$ are equal to the corresponding transfer entropy rates between $v$ and $f$. Therefore, the information transmission can be defined through the latter transfer entropy rates,
\begin{align}
\dot{T}_{x\to f}=\dot{T}_{v\to f}&=\lim_{\delta t\to0+}\frac{1}{\delta t}\left\langle\log\frac{P[f(t+\delta t)|\mathbf{f},\mathbf{v}]}{P[f(t+\delta t)|\mathbf{f}]}\right\rangle , \mathcal{T}_{\mathrm{FF}}=\dot{T}_{v\to f}/H,\label{eqs:spatialtvf}\\
\dot{T}_{f\to x}=\dot{T}_{f\to v}&=\lim_{\delta t\to0+}\frac{1}{\delta t}\left\langle\log\frac{P[v(t+\delta t)|\mathbf{f},\mathbf{v}]}{P[v(t+\delta t)|\mathbf{v}]}\right\rangle , \mathcal{T}_{\mathrm{FB}}=\dot{T}_{f\to v}/F,\label{eqs:spatialtfv}
\end{align}
where $\mathbf{v}$ and $\mathbf{f}$ are abbreviations for the steady-state trajectories up to time $t$, $\mathbf{v}_{[-\infty,t]}$ and $\mathbf{f}_{[-\infty,t]}$, the logarithms are to base $e$, the unit of $\mathcal{T}_{\mathrm{FF}}$ and $\mathcal{T}_{\mathrm{FB}}$ is nats, and the angular brackets $\langle\cdot\rangle$ denote an average over the temporal trajectories in steady state. The navigation performance is defined as the peak localization,
\begin{align}
\mathcal{P}= 1/\sigma_{x}^{2}\;,\;\mathcal{P}_{0}=1/\Delta x_{0}^{2}=H^{2}/\sigma_{v0}^{2},
\end{align}
where $\sigma_{x}$ is the standard deviation of $x$ in steady-state, and $\Delta x_{0}$ and $\sigma_{v0}$ are, respectively, the persistence length and the standard deviation of velocity in steady-state in the absence of actuation, i.e., when $J=0$.

\subsection*{Transfer entropy rates from conditional variance} We exploit the linearity of the model to derive the transfer entropy rates. Since the drift terms in Eqs. \ref{eqs:spatialx}-\ref{eqs:spatialv} are linear in $x$, $f$ and $v$, and all the noise terms are Gaussian distributed, the joint distribution of the stochastic trajectories of $x$, $f$ and $v$ in a steady-state is Gaussian. Therefore, all marginal and conditional distributions of the trajectories are Gaussian as well. Specifically, the probability distributions appearing in the numerators and denominators of Eqs. \ref{eqs:spatialtvf} and \ref{eqs:spatialtfv} are Gaussian. This simplifies the transfer entropy rates,
\begin{align}
\dot{\mathcal{T}}_{v\to f}=\lim_{\delta t\to 0+}\frac{1}{\delta t}\Biggl[ &\int D[\mathbf{f}]\;D[\mathbf{v}]\;P[\mathbf{f},\mathbf{v}]\int D[f(t+\delta t)]\;P[f(t+\delta t)|\mathbf{f},\mathbf{v}]\log P[f(t+\delta t)|\mathbf{f},\mathbf{v}]\nonumber\\
-&\int D[\mathbf{f}]\;D[\mathbf{v}]\; P[\mathbf{f},\mathbf{v}]\int D[f(t+\delta t)]\;P[f(t+\delta t)|\mathbf{f}]\log P[f(t+\delta t)|\mathbf{f}] \Biggr] \\
=\lim_{\delta t\to 0+}\frac{1}{\delta t}\Biggl[ &-\frac{1}{2}\log \sigma^{2}_{f(t+\delta t)|\mathbf{f},\mathbf{v}}+\frac{1}{2}\log\sigma^{2}_{f(t+\delta t)|\mathbf{f}}\Biggr] =\lim_{\delta t\to 0+}\frac{1}{2\delta t}\log\frac{\sigma^{2}_{f(t+\delta t)|\mathbf{f}}}{\sigma^{2}_{f(t+\delta t)|\mathbf{f},\mathbf{v}}},\label{eqs:spatialsigmavf}\\
\mathrm{and~}\dot{\mathcal{T}}_{f\to v}=\lim_{\delta t\to 0+}\frac{1}{2\delta t}&\log\frac{\sigma^{2}_{v(t+\delta t)|\mathbf{v}}}{\sigma^{2}_{v(t+\delta t)|\mathbf{f},\mathbf{v}}}\mathrm{,~analogously}\label{eqs:spatialsigmafv}.
\end{align}

The conditional variances in the above expressions are not, in general, directly available from the equations of motion of the model. For example, $\sigma^{2}_{f(t+\delta t)|\mathbf{f}}$ is the variance of $f(t+\delta t)$ when only the trajectory $\mathbf{f}$ is known. However, the equation of motion of $f$, Eq. \ref{eqs:spatialf}, states that the future value of $f$ depends on the current values of both $f$ and $x$. The current value of $x$ is not directly known from the conditioning on the trajectory $\mathbf{f}$. So the conditional variance $\sigma^{2}_{f(t+\delta t)|\mathbf{f}}$ cannot be directly obtained from the equation of motion. However, using stochastic control theory, we can derive the conditional variances from a set of causal functions, as shown next.

\subsection*{Conditional variance from causal functions}
For each conditional variance, we can find an \textit{effective} equation of motion that depends on only the conditioned trajectory and is statistically consistent with the original equation of motion \cite{chetrite2019information}. For example, the \textit{effective dynamics} of $f$ within the conditioned space of $\mathbf{f}$ is defined such that it reproduces the distribution $P[f(t+\delta t)|\mathbf{f}]$ that originally emerged from Eqs. \ref{eqs:spatialx}-\ref{eqs:spatialv}. This effective dynamics is defined through, in general, a non-Markovian \textit{causal function} $H^{(f)}(t)$ that satisfies the linear equation
\begin{align}\label{eqs:fcausal}
f(t)=\int_{-\infty}^{t}ds~H^{(f)}(t-s)\xi^{(f)}(s),
\end{align}
where $\xi^{(f)}(t)$ is a Gaussian white noise with variance $2D_{f}$. Since the trajectory of $f$ and $\xi^{(f)}$ map to each other one-to-one, conditioning on $\mathbf{f}$ is equivalent to knowing the past trajectory of the noise $\xi^{(f)}$, denoted as $\boldsymbol{\xi}^{(f)}$. Therefore, the conditional variance $\sigma^{2}_{f(t+\delta t)|\mathbf{f}}=\sigma^{2}_{f(t+\delta t)|\boldsymbol{\xi}^{(f)}}$. The latter is obtained as the conditional variance of Eq. \ref{eqs:fcausal}, as shown below.

The conditional variance is obtained from the first and second conditional moments of Eq. \ref{eqs:fcausal},
\begin{align}
\sigma^{2}_{f(t+\delta t)|\mathbf{f}}=\sigma^{2}_{f(t+\delta t)|\boldsymbol{\xi}^{(f)}}=\Bigl\langle f(t+\delta t)^{2}-\langle f(t+\delta t)|\boldsymbol{\xi}^{(f)}\rangle^{2}\Bigr\rangle _{\boldsymbol{\xi}^{(f)}},\label{eqs:condvarf}
\end{align}
where we have emphasized with a subscript that the outermost angular brackets denote an average over $\boldsymbol{\xi}^{(f)}$. We use Eq. \ref{eqs:fcausal} on each of the two terms. The first conditional moment becomes,
\begin{align}
\langle f(t+\delta t)|\boldsymbol{\xi}^{(f)}\rangle&=\left\langle \int_{-\infty}^{t+\delta t} ds \;H^{(f)}(t+\delta t-s)\xi^{(f)}(s) \Big|\boldsymbol{\xi}^{(f)} \right\rangle\\
&=\int_{-\infty}^{t} ds \;H^{(f)}(t+\delta t-s)\left\langle \xi^{(f)}(s) \Big|\boldsymbol{\xi}^{(f)} \right\rangle+\int_{t}^{t+\delta t} ds \;H^{(f)}(t+\delta t-s)\left\langle \xi^{(f)}(s) \Big|\boldsymbol{\xi}^{(f)} \right\rangle\\
&=\int_{-\infty}^{t} ds \;H^{(f)}(t+\delta t-s) \xi^{(f)}(s) +0,
\end{align}
where we used the fact that the conditioned trajectory fully specifies the past noise and does not specify the future noise. Squaring and averaging gives
\begin{align}
\Bigl\langle\langle f(t+\delta t)|\boldsymbol{\xi}^{(f)}\rangle^{2}\Bigr\rangle _{\boldsymbol{\xi}^{(f)}}&=\int_{-\infty}^{t} ds \int_{-\infty}^{t} ds^{'}\;H^{(f)}(t+\delta t-s)H^{(f)}(t+\delta t-s^{'}) \langle\xi^{(f)}(s)\xi^{(f)}(s^{'})\rangle\\
&=\int_{-\infty}^{t} ds \int_{-\infty}^{t} ds^{'}\;H^{(f)}(t+\delta t-s)H^{(f)}(t+\delta t-s^{'}) \;2D_{f}\delta(s-s^{'})\\
&=2D_{f}\int_{-\infty}^{t} ds \;H^{(f)}(t+\delta t-s)^{2}\label{eqs:ffirst}.
\end{align}
Similarly, using Eq. \ref{eqs:fcausal}, the second moment of $f(t+\delta t)$ is
\begin{align}
\Bigl\langle f(t+\delta t)^{2}\Bigr\rangle _{\boldsymbol{\xi}^{(f)}}&=\int_{-\infty}^{t+\delta t} ds \int_{-\infty}^{t+\delta t} ds^{'}\;H^{(f)}(t+\delta t-s)H^{(f)}(t+\delta t-s^{'}) \langle\xi^{(f)}(s)\xi^{(f)}(s^{'})\rangle\\
&=\int_{-\infty}^{t+\delta t} ds \int_{-\infty}^{t+\delta t} ds^{'}\;H^{(f)}(t+\delta t-s)H^{(f)}(t+\delta t-s^{'}) \;2D_{f}\delta(s-s^{'})\\
&=2D_{f}\int_{-\infty}^{t+\delta t} ds \;H^{(f)}(t+\delta t-s)^{2}\label{eqs:fsecond}.
\end{align}
Subtracting Eq. \ref{eqs:ffirst} from \ref{eqs:fsecond} gives the conditional variance in Eq. \ref{eqs:condvarf},
\begin{align}
\sigma^{2}_{f(t+\delta t)|\mathbf{f}}=2D_{f}\int_{t}^{t+\delta t}ds\;H^{(f)}(t+\delta t -s)^{2}=2D_{f}\int_{0}^{\delta t}du\;H^{(f)}(u)^{2},\label{eqs:fsigmacausal}
\end{align}
where in the last step we substituted $u=t+\delta t-s$. This expression gives the conditional variance $\sigma^{2}_{f(t+\delta t)|\mathbf{f}}$ in terms of the causal function $H^{(f)}(t)$.

The above expression naturally motivates the choice of a boundary condition for the causal function in order for the diffusion constant to be consistent with the original equation of motion, Eq. \ref{eqs:spatialf}. 
In the limit of small $\delta t$, both Eqs. \ref{eqs:spatialf} and \ref{eqs:fsigmacausal} must yield the same diffusion constant. The diffusion constant of $f$ in the original dynamics is $2D_{f}$.
The diffusion constant from Eq. \ref{eqs:fsigmacausal} is $\lim_{\delta t \to 0+}\sigma_{f(t+\delta t)|\mathbf{f}}^{2}/\delta t=2D_{f}H^{(f)}(0+)^{2}$. In order for them to be the same, $H^{(f)}(0+)^{2}=1$. This fixes only the magnitude of $H^{(f)}(0+)$ to be 1, not the phase. However, we impose the additional \textit{minimum-phase condition} from control theory, which ensures that the trajectories $f(t)$ and $\xi^{(f)}(t)$ remain mappable one-to-one \cite{chetrite2019information,rosinberg2018influence}. Therefore, we \textit{choose} to define the boundary condition $H^{(f)}(0+)=1$.

The causal functions corresponding to the other conditional variances in Eqs. \ref{eqs:spatialsigmavf} and \ref{eqs:spatialsigmafv} are defined analogously,
\begin{align}
v(t)&=\int_{-\infty}^{t}ds~H^{(v)}(t-s)\xi^{(v)}(s)\label{eqs:vcausal},\\
\begin{bmatrix}f(t)\\v(t)
\end{bmatrix} &=\int_{-\infty}^{t}ds\;\mathbf{H}^{(fv)}(t-s)
\begin{bmatrix}\zeta^{(f)}(s)\\ \zeta^{(v)}(s)
\end{bmatrix}=\int_{-\infty}^{t}ds 
\begin{bmatrix}H^{(fv)}_{ff}(t-s)\zeta^{(f)}(s)+H^{(fv)}_{fv}(t-s)\zeta^{(v)}(s)
\\ H^{(fv)}_{vf}(t-s)\zeta^{(f)}(s)+H^{(fv)}_{vv}(t-s)\zeta^{(v)}(s)
\end{bmatrix} ,\label{eqs:fvcausal}
\end{align}
where $\xi^{(v)}(t)$ is a Gaussian white noise with variance $2D_{v}$, and $\begin{bmatrix}\zeta^{(f)}(t)\\ \zeta^{(v)}(t)\end{bmatrix}$ are Gaussian white noise with covariance matrix
$\mathbf{D}^{(fv)}=\begin{pmatrix}2D_{f}&0\\0&2D_{v}\end{pmatrix}$.
Proceeding similarly to the variance of Eq. \ref{eqs:fcausal}, the variance of Eq. \ref{eqs:vcausal} will give $\sigma^{2}_{v(t+\delta t)|\mathbf{v}}$ in terms of the corresponding causal function,
\begin{align}
\sigma^{2}_{v(t+\delta t)|\mathbf{v}}=2D_{v}\int_{0}^{\delta t}du\;H^{(v)}(u)^{2},\label{eqs:vsigmacausal}
\end{align}
and the variance of Eq. \ref{eqs:fvcausal} gives $\sigma^{2}_{f(t+\delta t)|\mathbf{f},\mathbf{v}}$ and $\sigma^{2}_{v(t+\delta t)|\mathbf{f},\mathbf{v}}$ in terms of the corresponding causal functions,
\begin{align}
\sigma_{f(t+\delta t)|\mathbf{f},\mathbf{v}}^{2}&=2D_{f}\int_{0}^{\delta t}du~H^{(fv)}_{ff}(u)^{2}+2D_{v}\int_{0}^{\delta t}du~H^{(fv)}_{fv}(u)^{2}\label{eqs:fvsigmacausal},\\
\sigma_{v(t+\delta t)|\mathbf{f},\mathbf{v}}^{2}&=2D_{f}\int_{0}^{\delta t}du~H^{(fv)}_{vf}(u)^{2}+2D_{v}\int_{0}^{\delta t}du~H^{(fv)}_{vv}(u)^{2}\label{eqs:vfsigmacausal}.
\end{align}
The boundary conditions for these causal functions are also defined as motivated by the minimum phase condition, $H^{(v)}(0+)=H^{(fv)}_{ff}(0+)=H^{(fv)}_{vv}(0+)=1$, and $H^{(fv)}_{fv}(0+)=H^{(fv)}_{vf}(0+)=0$. 

\subsection*{Transfer entropy rates directly from causal functions}
We can, now, directly express the transfer entropy rates in Eqs. \ref{eqs:spatialsigmavf} and \ref{eqs:spatialsigmafv} in terms of the causal functions. For the feedforward transfer entropy rate $\dot{\mathcal{T}}_{v\to f}$, we substitute Eqs. \ref{eqs:fsigmacausal} and \ref{eqs:fvsigmacausal} in Eq. \ref{eqs:spatialsigmavf},
\begin{align}
\dot{\mathcal{T}}_{v\to f}&=\lim_{\delta t\to 0+}\frac{1}{2\delta t} \log\frac{2D_{f}\int_{0}^{\delta t}du~H^{(f)}(u)^{2}}{2D_{f}\int_{0}^{\delta t}du~H^{(fv)}_{ff}(u)^{2}+2D_{v}\int_{0}^{\delta t}du~H^{(fv)}_{fv}(u)^{2}}.
\end{align}
In order to evaluate this expression in the limit $\delta t \to 0$, we can simplify the expression by Taylor expanding the numerator and denominator of the logarithm for small $\delta t$,
\begin{align}
\dot{\mathcal{T}}_{v\to f}&=\lim_{\delta t\to 0+}\frac{1}{2\delta t} \log\Bigl[ \{2D_{f}\delta t H^{(f)}(0+)^{2}+2D_{f}\delta t^{2}H^{(f)}(0+)\dot{H}^{(f)}(0+)\}\nonumber\\
& /\{2D_{f}\delta t H^{(fv)}_{ff}(0+)^{2}+2D_{f}\delta t^{2}H^{(fv)}_{ff}(0+)\dot{H}^{(fv)}_{ff}(0+)+2D_{v}\delta t H^{(fv)}_{fv}(0+)^{2}+2D_{v}\delta t^{2}H^{(fv)}_{fv}(0+)\dot{H}^{(fv)}_{fv}(0+)\}\Bigr]
\end{align}
This can be simplified further using the boundary conditions that were defined under Eqs. \ref{eqs:fsigmacausal} and \ref{eqs:vfsigmacausal}, $H^{(f)}(0+)=H^{(fv)}_{ff}(0+)=1$, and $H^{(fv)}_{fv}(0+)=0$. This gives,
\begin{align}
\dot{\mathcal{T}}_{v\to f}
&=\lim_{\delta t\to 0+}\frac{1}{2\delta t} \log\frac{2D_{f}\delta t+2D_{f}\delta t^{2}\dot{H}^{(f)}(0+)}{2D_{f}\delta t+2D_{f}\delta t^{2}\dot{H}^{(fv)}_{vv}(0+)}\\
&=\lim_{\delta t\to 0+}\frac{1}{2\delta t} \log\frac{1+\delta t \dot{H}^{(f)}(0+)}{1+\delta t\dot{H}^{(fv)}_{ff}(0+)}\\
&=\lim_{\delta t\to 0+}\frac{1}{2\delta t} \Bigl[ \delta t\dot{H}^{(f)}(0+)-\delta t\dot{H}^{(fv)}_{ff}(0+)\Bigr]\\
&=\frac{1}{2}\Bigl[ \dot{H}^{(f)}(0+)-\dot{H}^{(fv)}_{ff}(0+)\Bigr] \label{eqs:tvfcausal}.
\end{align}
An analogous derivation gives the feedback transfer entropy rate as
\begin{align}
\dot{\mathcal{T}}_{f\to v}=\frac{1}{2}\Bigl[ \dot{H}^{(v)}(0+)-\dot{H}^{(fv)}_{vv}(0+)\Bigr] .\label{eqs:tfvcausal}
\end{align}
Therefore, if we can obtain analytical expressions for the causal functions, we can derive the transfer entropy rates.

\subsection*{Solving for causal functions in frequency space} The causal functions for the effective dynamics can be solved by imposing full consistency with the original dynamics. For analytical convenience, we impose this consistency in the frequency space.

We start by defining our convention for Fourier transforms. The Fourier and inverse Fourier transforms for any function $h(t)$ and its frequency-space representation $\hat{h}(\omega)$ are
\begin{align}
\hat{h}(\omega)=\int_{-\infty}^{\infty} dt\;e^{i\omega t}h(t),\qquad h(t)=\frac{1}{2\pi}\int_{-\infty}^{\infty} d\omega\;e^{-i\omega t}\hat{h}(\omega).
\end{align}
We will sometimes eliminate the hat ($~\hat{}~$) whenever it is clear from the arguments of the function that it is in Fourier space.

\subsubsection*{Causal functions for $f(t)$} The consistency of the effective equation of motion for $f(t)$, Eq. \ref{eqs:fcausal}, with the original equation of motion, Eq. \ref{eqs:spatialf}, implies that the power spectra derived from both have to be the same. The power spectrum matrix can be derived using the linearity of the original equations of motion, $\mathbf{S}(\omega)=(\mathbf{A}-i\omega\mathbbm{1})^{-1}\cdot 2\mathbf{D}\cdot(\mathbf{A}^{T}+i\omega\mathbbm{1})^{-1}$, where $A=\begin{pmatrix}0&0&-1\\Gk&F&0\\0&-J&H\end{pmatrix}$ and $D=\begin{pmatrix}0&0&0\\0&D_{f}&0\\0&0&D_{v}\end{pmatrix}$ \cite{gardiner1985handbook}. This gives the power spectrum of $f(t)$ as
\begin{align}
S_{ff}(\omega)=2D_{f}\frac{D_v G^2 k^2/D_f + \omega^2 H^2 + \omega^4}{G^2 k^2 J^2 +[F^{2}H^{2}- 2 G k (F + H) J ]\omega^2 + (F^{2}+H^{2})\omega^{4}+\omega^{6}}.
\end{align}
On the other hand, we can derive the power spectrum from the effective equation of motion, Eq. \ref{eqs:fcausal}, as
\begin{align}
S_{ff}(\omega)=\langle \hat{f}(\omega)\hat{f}(-\omega)\rangle=H^{(f)}(\omega)\langle\hat{\xi}^{(f)}(\omega)\hat{\xi}^{(f)}(-\omega)\rangle H^{(f)}(-\omega)=H^{(f)}(\omega)\cdot 2D_{f}\cdot H^{(f)}(-\omega).
\end{align}
Equating the above two expressions, we get a quadratic equation for $H^{(f)}(\omega)$,
\begin{align}
H^{(f)}(\omega)\cdot H^{(f)}(-\omega)=\frac{D_v G^2 k^2/D_f + \omega^2 H^2 + \omega^4}{G^2 k^2 J^2 +[F^{2}H^{2}- 2 G k (F + H) J ]\omega^2 + (F^{2}+H^{2})\omega^{4}+\omega^{6}}.
\end{align}

For solving this equation, we perform a Wiener-Hopf factorization of the power spectrum \cite{chetrite2019information}. We assume that the causal function satisfies the ansatz
\begin{align}
H^{(f)}(\omega)=\frac{i(\omega-\alpha_{1})(\omega-\beta_{1})}{(\omega-\gamma_{1})(\omega-\delta_{1})(\omega-\epsilon_{1})},
\end{align}
where $\alpha_{1},\beta_{1},\gamma_{1},\delta_{1}$ and $\epsilon_{1}$ are complex functions of the system parameters. We \textit{choose} $(\alpha_{1},\beta_{1},\dots,\epsilon_{1})$ such that they all have a negative imaginary part. This is equivalent to the function $H^{(f)}(\omega)$ being both pole-free and zero-free in the upper complex plane. The absence of poles ensures that $H^{(f)}(t<0)=0$, i.e., causality of the effective dynamics. The absence of zeros ensures that $1/H^{(f)}(\omega)$ is also causal, i.e., the equation $\hat{f}(\omega)=H^{(f)}(\omega)\hat{\xi}^{(f)}(\omega)$ can be uniquely inverted into $\hat{\xi}^{(f)}(\omega)=\hat{f}(\omega)/H^{(f)}(\omega)$, so that the trajectories $f(t)$ and $\xi^{(f)}(t)$ are mappable one-to-one, following the \textit{minimum phase condition} in control theory\cite{chetrite2019information}. We also choose the overall sign of the causal function such that it is positive in real space. 

Using this ansatz, $(\alpha_{1},\beta_{1},\dots,\epsilon_{1})$ can be obtained by equating the two expressions for the power spectrum,
\begin{align}
H^{(f)}(\omega)\cdot H^{(f)}(-\omega)=\frac{(\omega^{2}-\alpha_{1}^{2})(\omega^{2}-\beta_{1}^{2})}{(\omega^{2}-\gamma_{1}^{2})(\omega^{2}-\delta_{1}^{2})(\omega^{2}-\epsilon_{1}^{2})}=\frac{D_v G^2 k^2/D_f + \omega^2 H^2 + \omega^4}{G^2 k^2 J^2 +[F^{2}H^{2}- 2 G k (F + H) J ]\omega^2 + (F^{2}+H^{2})\omega^{4}+\omega^{6}},
\end{align}
where the coefficients of each power of $\omega$ in the numerator and denominator need to be matched. Specifically, $\alpha_{1}^{2}$ and $\beta_{1}^{2}$ are the roots of the quadratic equation
\begin{align}\label{eqs:Omega1}
\Omega^{2}+H^{2}\Omega+D_{v}G^{2}k^{2}/D_{f}=0,
\end{align}
where $\Omega=\omega^{2}$. Similarly, $\gamma_{1}^{2}$, $\delta_{1}^{2}$ and $\epsilon_{1}^{2}$ are the roots of the cubic equation
\begin{align}\label{eqs:Omega2}
\Omega^{3}+(F^{2}+H^{2})\Omega^{2}+[F^{2}H^{2}-2Gk(F+H)J]\Omega+G^{2}k^{2}J^{2}=0.
\end{align}
Therefore, $(\alpha_{1}^{2},\beta_{1}^{2},\dots,\epsilon_{1}^{2})$ satisfy the equations
\begin{align}
\alpha^{2}_{1}+\beta_{1}^{2}&=-H^{2},\label{eqs:rooteq1a}\\
\alpha_{1}^{2}\beta_{1}^{2}&=D_{v}G^{2}k^{2}/D_{f},\label{eqs:rooteq1b}\\
\gamma_{1}^{2}+\delta_{1}^{2}+\epsilon_{1}^{2}&=-(F^{2}+H^{2}),\label{eqs:rooteq1c}\\
\gamma_{1}^{2}\delta_{1}^{2}+\gamma_{1}^{2}\epsilon_{1}^{2}+\delta_{1}^{2}\epsilon_{1}^{2}&=F^{2}H^{2}-2Gk(F+H)J,\label{eqs:rooteq1d}\\
\gamma_{1}^{2}\delta_{1}^{2}\epsilon_{1}^{2}&=-G^{2}k^{2}J^{2}.\label{eqs:rooteq1e}
\end{align}

We do not need to fully solve the above equations to obtain the transfer entropy rate. The transfer entropy rate in Eq. \ref{eqs:tvfcausal} depends on only $\dot{H}^{(f)}(0+)$, not on the full time-dependent function $H^{(f)}(t)$. To obtain $\dot{H}^{(f)}(0+)$, and how it depends on $\alpha_{1},\beta_{1},\gamma_{1},\delta_{1}$ and $\epsilon_{1}$, we manipulate the inverse Fourier transform:
\begin{align}
H^{(f)}(t)&=\frac{1}{2\pi}\int_{-\infty}^{\infty}d\omega\; e^{-i\omega t}H^{(f)}(\omega),\\
&=\frac{1}{2\pi}\int_{-\infty}^{\infty}d\omega\; e^{-i\omega t} \frac{i(\omega-\alpha_{1})(\omega-\beta_{1})}{(\omega-\gamma_{1})(\omega-\delta_{1})(\omega-\epsilon_{1})}.\\
\implies\dot{H}^{(f)}(t)&=\frac{1}{2\pi}\int_{-\infty}^{\infty}d\omega\; (-i\omega) e^{-i\omega t} \frac{i(\omega-\alpha_{1})(\omega-\beta_{1})}{(\omega-\gamma_{1})(\omega-\delta_{1})(\omega-\epsilon_{1})}.\\
\implies\dot{H}^{(f)}(0+)&=\lim_{t\to 0+}\frac{1}{2\pi}\int_{-\infty}^{\infty}d\omega\; (-i\omega) e^{-i\omega t} \frac{i(\omega-\alpha_{1})(\omega-\beta_{1})}{(\omega-\gamma_{1})(\omega-\delta_{1})(\omega-\epsilon_{1})},\\
&=-i(\gamma_{1}+\delta_{1}+\epsilon_{1}-\alpha_{1}-\beta_{1}),
\end{align}
where, in going to the last line, the integral is evaluated using the residue theorem.

Therefore, for the contribution to the transfer entropy rate, we only need to obtain $\gamma_{1}+\delta_{1}+\epsilon_{1}-\alpha_{1}-\beta_{1}$ from Eqs. \ref{eqs:rooteq1a}-\ref{eqs:rooteq1e}. From the first two equations, we obtain,
\begin{align}
\alpha_{1}+\beta_{1}=[\alpha_{1}^{2}+\beta_{1}^{2}+2\alpha_{1}\beta_{1}]^{1/2}=[\alpha_{1}^{2}+\beta_{1}^{2}+2[\alpha_{1}^{2}\beta_{1}^{2}]^{1/2}]^{1/2}.
\end{align}
For deciding the sign of each square root, 
we appeal to the geometric form of the quadratic Eq. \ref{eqs:Omega1} in the limit where $G$ and $J$ are vanishingly small. The above analytical expression is expected to be valid for all $G$ and $J$, so it must be commensurate with this limit. When $G=0$, Eq. \ref{eqs:Omega1} has roots $\Omega=0,-H^{2}$. When $G$ is small but positive, the LHS of Eq. \ref{eqs:Omega1} evaluated at $\Omega=0$ and $\Omega=-H^{2}$ is positive in both cases, which implies that the roots have moved to $\Omega=-\delta_{a}$ and $\Omega=-H^{2}+\delta_{b}$ where $\delta_{a},\delta_{b}>0$. They remain purely real and negative. Hence, $\alpha_{1}$ and $\beta_{1}$, which are the square roots of those roots, are both purely imaginary, so should be taken to be negative imaginary according to our sign convention based on the \textit{minimum phase condition}. Therefore, $\alpha_{1}\beta_{1}$ should be negative; so, from Eq. \ref{eqs:rooteq1b}, it is equal to $-\sqrt{D_{v}G^{2}k^{2}/D_{f}}$. This implies,
\begin{align}
\alpha_{1}+\beta_{1}=\sqrt{-H^{2}-2\sqrt{D_{v}G^{2}k^{2}/D_{f}}}=-i\sqrt{H^{2}+2Gk\sqrt{D_{v}/D_{f}}},
\end{align}
where we have again taken the sum of $\alpha_{1}$ and $\beta_{1}$ to have a negative imaginary part, and each radical sign now denotes the principal value of the square root.

We proceed analogously for $\gamma_{1}+\delta_{1}+\epsilon_{1}$. We can relate this sum to Eqs. \ref{eqs:rooteq1c}-\ref{eqs:rooteq1e} as
\begin{align}
I\equiv\gamma_{1}+\delta_{1}+\epsilon_{1}&=[\gamma_{1}^{2}+\delta_{1}^{2}+\epsilon_{1}^{2}+2\gamma_{1}\delta_{1}+2\delta_{1}\epsilon_{1}+2\gamma_{1}\epsilon_{1}]^{1/2}\\
&=\left[ \gamma_{1}^{2}+\delta_{1}^{2}+\epsilon_{1}^{2}+2[\gamma_{1}^{2}\delta_{1}^{2}+\delta_{1}^{2}\epsilon_{1}^{2}+\gamma_{1}^{2}\epsilon_{1}^{2}+2\gamma_{1}\delta_{1}^{2}\epsilon_{1}+2\gamma_{1}\delta_{1}\epsilon_{1}^{2}+2\gamma_{1}^{2}\delta_{1}\epsilon_{1}]^{1/2}\right] ^{1/2}\\
&=\left[ \gamma_{1}^{2}+\delta_{1}^{2}+\epsilon_{1}^{2}+2[\gamma_{1}^{2}\delta_{1}^{2}+\delta_{1}^{2}\epsilon_{1}^{2}+\gamma_{1}^{2}\epsilon_{1}^{2}+2\gamma_{1}\delta_{1}\epsilon_{1}(\gamma_{1}+\delta_{1}+\epsilon_{1})]^{1/2}\right] ^{1/2}\\
&=\left[ \gamma_{1}^{2}+\delta_{1}^{2}+\epsilon_{1}^{2}+2\left[ \gamma_{1}^{2}\delta_{1}^{2}+\delta_{1}^{2}\epsilon_{1}^{2}+\gamma_{1}^{2}\epsilon_{1}^{2}+2\gamma_{1}\delta_{1}\epsilon _{1}I\right] ^{1/2}\right] ^{1/2}\label{eqs:selfcons1},
\end{align}
which is a self-consistent equation in $I$. 

We first determine the correct signs for each square root. For this, we appeal again to the expression being commensurate with the limit where the gains $G$ and $J$ are vanishingly small. When $G=J=0$, Eq. \ref{eqs:Omega2} has solutions $\Omega=0,-F^{2},-H^{2}$. When $G$ and $J$ are small but positive, the LHS of Eq. \ref{eqs:Omega2}, evaluated at $\Omega=0$, $\Omega=-F^{2}$ and $\Omega=-H^{2}$, is positive in all cases. Combined with the fact that the coefficient of $\Omega^{3}$ is positive, this implies that the roots become either $-\delta_{c},-F^{2}+\delta_{d},-H^{2}-\delta_{e}$ if $F<H$, or $-\delta_{c},-H^{2}+\delta_{d},-F^{2}-\delta_{e}$ if $F>H$, for $\delta_{c},\delta_{d},\delta_{e}>0$. In both cases, $\gamma_{1}$, $\delta_{1}$ and $\epsilon_{1}$, which are the square roots of these roots, are purely imaginary, and so should be taken to be negative imaginary according to our sign convention. This means that $\gamma_{1}+\delta_{1}+\epsilon_{1}$ is negative imaginary, $(\gamma_{1}+\delta_{1}+\epsilon_{1})^{2}$ is negative real, $\gamma_{1}\delta_{1}+\delta_{1}\epsilon_{1}+\gamma_{1}\epsilon_{1}$ is negative real, $(\gamma_{1}\delta_{1}+\delta_{1}\epsilon_{1}+\gamma_{1}\epsilon_{1})^{2}$ is positive real, $\gamma_{1}\delta_{1}\epsilon_{1}$ is positive imaginary, and $\gamma_{1}^{2}\delta_{1}^{2}\epsilon_{1}^{2}$ is negative real. Now, applying Eqs. \ref{eqs:rooteq1c}-\ref{eqs:rooteq1e} gives,
\begin{align}
I=-\sqrt{-(F^{2}+H^{2})-2\sqrt{F^{2}H^{2}-2GkJ(F+H)+2\sqrt{-G^{2}k^{2}J^{2}}I}}\label{eqs:selfcons2},
\end{align}
where each square root now denotes the principal value.

The solution to the self-consistent equation is,
\begin{align}
I=-i(F+H),
\end{align}
which can be verified by substitution.

Therefore, we have,
\begin{align}
\dot{H}^{(f)}(0+)=-i(\gamma_{1}+\delta_{1}+\epsilon_{1}-\alpha_{1}-\beta_{1})&=-i\left[ -i(F+H)+i\sqrt{H^{2}+2Gk\sqrt{D_{v}/D_{f}}}\right] \\
&=\sqrt{H^{2}+2Gk\sqrt{D_{v}/D_{f}}}-F-H.\label{eqs:tvf1term}
\end{align}
For the transfer entropy rate $\dot{\mathcal{T}}_{v\to f}$ in Eq. \ref{eqs:tvfcausal}, we also need $\dot{H}^{(fv)}_{ff}(0+)$. For this we need to first obtain the causal function matrix $\mathbf{H}^{(fv)}(t)$ that was defined in Eq. \ref{eqs:fvcausal}. However, the only sources of noise in the system is in the equations of motion of $f$ and $v$, the two-component causal function $\mathbf{H}^{(fv)}(t)$ is simply a sub-matrix of the three-component causal function obtained from the original equations of motion, Eqs. \ref{eqs:spatialx}-\ref{eqs:spatialv}. This can be seen by integrating the original equations of motion, which yields, in a steady-state,
\begin{align}
\begin{bmatrix}x(t)\\f(t)\\v(t)\end{bmatrix}&=\int_{-\infty}^{t}ds\; e^{-\mathbf{A}(t-s)}\begin{bmatrix}0\\\sqrt{2D_{f}}\psi(s)\\\sqrt{2D_{v}}\eta(s)\end{bmatrix},
\end{align}
where $\mathbf{A}=\begin{pmatrix}0&0&-1\\Gk&F&0\\0&-J&H\end{pmatrix}$ is the spring-constant matrix. Comparing this with Eq. \ref{eqs:fvcausal} shows that $\mathbf{H}^{(fv)}(t)$ is the sub-matrix of $e^{-\mathbf{A}t}$ in the space of $f$ and $v$. Therefore, $\dot{\mathbf{H}}^{(fv)}(t)$ is the corresponding sub-matrix of $-\mathbf{A}e^{-\mathbf{A}t}$, and $\dot{H}^{(fv)}_{ff}(0+)=-A_{ff}=-F$.

Substituting this and Eq. \ref{eqs:tvf1term} in Eq. \ref{eqs:tvfcausal}, we find the feedforward transfer entropy rate as
\begin{align}
\dot{\mathcal{T}}_{v\to f}=\frac{1}{2}\left[ \sqrt{H^{2}+2Gk\sqrt{D_{v}/D_{f}}}-H\right].
\label{eq:Td_vf}
\end{align}

\subsubsection*{Causal functions for $v(t)$} To obtain the feedback transfer entropy, we need the causal functions for $v(t)$. They are derived by analogy with those in the previous section. First, we impose consistency of the effective equation of motion for $v(t)$, Eq. \ref{eqs:vcausal}, in terms of the original equation of motion, Eq. \ref{eqs:spatialv}.
For this we equate the power spectrum of $v(t)$ as obtained from both equations of motion, in frequency space. From the original equation of motion, the power spectrum is derived as before,
\begin{align}
S_{vv}(\omega)=2D_{v}\frac{(F^{2}+D_{f}J^{2}/D_{v})\omega^{2} + \omega^4}{G^2 k^2 J^2 +[F^{2}H^{2}- 2 G k (F + H) J ]\omega^2 + (F^{2}+H^{2})\omega^{4}+\omega^{6}}.
\end{align}
From the effective equation of motion, the power spectrum is derived as $S_{vv}(\omega)=H^{(v)}(\omega)\cdot 2D_{v}\cdot H^{(v)}(-\omega)$. For solving for $H^{(v)}(\omega)$, we again propose a Wiener-Hopf factorization,
\begin{align}
H^{(v)}(\omega)=\frac{i(\omega-\alpha_{2})(\omega-\beta_{2})}{(\omega-\gamma_{1})(\omega-\delta_{1})(\omega-\epsilon_{1})},
\end{align}
where the denominator is identical to that of $H^{(f)}(\omega)$, since the denominator of $S_{vv}(\omega)$ is identical to that of $S_{ff}(\omega)$. The zeros $\alpha_{2}$ and $\beta_{2}$ are defined to have negative imaginary parts in line with the \textit{minimum phase condition}, similar to the previous section.

To solve for $\alpha_{2}$ and $\beta_{2}$, we equate the power spectra obtained from the effective and the original equations of motion. This gives
\begin{align}
\alpha_{2}^{2}+\beta_{2}^{2}&=-(F^{2}+D_{f}J^{2}/D_{v}),\\
\alpha_{2}^{2}\beta_{2}^{2}&=0.
\end{align}
These equations immediately give $\alpha_{2}+\beta_{2}=[\alpha_{2}^{2}+\beta_{2}^{2}]^{1/2}=-i\sqrt{F^{2}+D_{f}J^{2}/D_{v}}$. Following the same strategy as in the previous section, we find
\begin{align}
\dot{H}^{(v)}(0+)=-i(\gamma_{1}+\delta_{1}+\epsilon_{1}-\alpha_{2}-\beta_{2})=\sqrt{F^{2}+D_{f}J^{2}/D_{v}}-F-H.
\end{align}
Additionally, the causal function $H^{(fv)}_{vv}(t)$, appearing in Eq. \ref{eqs:tfvcausal}, is simply an element of $e^{-\mathbf{A}t}$, analogous to the previous section. Therefore, $\dot{H}^{(fv)}_{vv}(0+)=-A_{vv}=-H$.

Therefore, the feedback transfer entropy rate is given by
\begin{align}
\dot{\mathcal{T}}_{f\to v}=\frac{1}{2}\left[ \sqrt{F^{2}+D_{f}J^{2}/D_{v}}-F\right].
\label{eq:Td_fv}
\end{align}

\subsection*{Performance}
The navigation performance is defined as the degree of localization of the cell in steady-state, given by the inverse of the positional variance, $\mathcal{P}=1/\sigma_{x}^{2}$. $\sigma_{x}^{2}$ is an element of the steady-state covariance matrix $\mathbf{C}$. $\mathbf{C}$ is obtained by solving the Lyapunov equation $\mathbf{AC}+\mathbf{CA}^{T}=2\mathbf{D}$, where $\mathbf{A}$ and $\mathbf{D}$ are the spring-constant and diffusion-constant matrices, respectively. Analytically solving this gives
\begin{align}\label{eqs:statcov}
\mathbf{C}=\begin{pmatrix}\frac{D_{v}[F^{2}(F+H)+GkJ]+D_{f}(F+H)J^{2}}{GkJ[FH(F+H)-GkJ]}&-\frac{D_{v}F(F+H)+D_{f}J^{2}}{J[FH(F+H)-GkJ]}&0\\-\frac{D_{v}F(F+H)+D_{f}J^{2}}{J[FH(F+H)-GkJ]}& \frac{(F+H)(D_{v}Gk+D_{f}JH)}{J[FH(F+H)-GkJ]}&\frac{D_{v}Gk+D_{f}JH}{[FH(F+H)-GkJ]}\\0&\frac{D_{v}Gk+D_{f}JH}{[FH(F+H)-GkJ]}&\frac{D_{v}F(F+H)+D_{f}J^{2}}{[FH(F+H)-GkJ]}\end{pmatrix}.
\end{align}
Therefore, the navigation performance is given by the inverse of the first diagonal element of $\mathbf{C}$,
\begin{align}
\mathcal{P}&=\frac{GkJ[FH(F+H)-GkJ]}{D_{f}(F+H)J^{2}+D_{v}[F^{2}(F+H)+GkJ]}.\label{eqs:fullP}
\end{align}

\newpage
\section*{BEST for spatial sensing}
The Behavioral Equations of State (BESTs) express the navigation performance solely in terms of information. In this section, we derive the BEST for spatial sensing systems and the resultant information-theoretic bounds on navigation performance.

\subsection*{Derivation of BEST} The feedforward and feedback information and the performance, as derived in the previous section (see Eqs. \ref{eq:Td_vf}, \ref{eq:Td_fv} and \ref{eqs:fullP}), are given by,
\begin{align}
\mathcal{T}_{\mathrm{FF}}&=\dot{\mathcal{T}}_{v\to f}/H=\frac{1}{2}\left[ \sqrt{1+\frac{2Gk}{H^{2}}\sqrt{\frac{D_{v}}{D_{f}}}}-1\right] =\frac{Gk}{2H^{2}}\sqrt{\frac{D_{v}}{D_{f}}}+\mathcal{O}(k^{2}),\label{eqs:tff}\\
\mathcal{T}_{\mathrm{FB}}&=\dot{\mathcal{T}}_{f\to v}/F=\frac{1}{2}\left[ \sqrt{1+\frac{J^{2}}{F^{2}}\cdot\frac{D_{f}}{D_{v}}}-1\right] ,\label{eqs:tfb}\\
\mathcal{P}/\mathcal{P}_{0}&=\frac{GkJD_{v}[FH(F+H)-GkJ]}{H^{3}\{D_{f}(F+H)J^{2}+D_{v}[F^{2}(F+H)+GkJ]\}}=\frac{GkD_{v}}{H^{3}\left[\frac{D_{f}J}{FH}+\frac{D_{v}F}{JH}\right] }+\mathcal{O}(k^{2})\label{eqs:fullperf},
\end{align}
where we have shown the full expressions as well as the approximations at weak signal strengths $k$.

The connection between the performance and the information becomes evident when the performance is rewritten in the form,
\begin{align}
\mathcal{P}/\mathcal{P}_{0}&=
\frac{\left( \frac{Gk}{H^{2}}\sqrt{\frac{D_{v}}{D_{f}}}\right)\left(\frac{J}{F}\sqrt{\frac{D_{f}}{D_{v}}}\right) \left[ \left( \frac{F}{H}\right) ^{2}+\left( \frac{F}{H}\right) - \left( \frac{F}{H}\right) \left( \frac{Gk}{H^{2}}\sqrt{\frac{D_{v}}{D_{f}}}\right)\left(\frac{J}{F}\sqrt{\frac{D_{f}}{D_{v}}}\right) \right] }
{\left( \frac{F}{H}\right)\left( 1+\left( \frac{F}{H}\right)\right) \left(\frac{J}{F}\sqrt{\frac{D_{f}}{D_{v}}}\right) ^{2}+\left( \frac{F}{H}\right)\left( 1+\left( \frac{F}{H}\right)\right) +\left( \frac{Gk}{H^{2}}\sqrt{\frac{D_{v}}{D_{f}}}\right)\left(\frac{J}{F}\sqrt{\frac{D_{f}}{D_{v}}}\right) }\\
&=
\frac{\left( \frac{Gk}{H^{2}}\sqrt{\frac{D_{v}}{D_{f}}}\right) }{\left( \frac{J}{F}\sqrt{\frac{D_{f}}{D_{v}}}\right) +\left(\frac{J}{F}\sqrt{\frac{D_{f}}{D_{v}}}\right) ^{-1}}+\mathcal{O}(k^{2})\label{eqs:perf-spatial-lin}.
\end{align}
Thus, the performance depends on only three collective variables of the model: $(Gk/H^{2})\sqrt{D_{v}/D_{f}}$, $(J/F)\sqrt{D_{f}/{D_{v}}}$, and $F/H$. The former two monotonically map onto the feedforward and feedback information in Eqs. \ref{eqs:tff} and \ref{eqs:tfb}, $(Gk/H^{2})\sqrt{D_{v}/D_{f}}\equiv\widetilde{\mathcal{T}}_{\mathrm{FF}}=[(2\mathcal{T}_{\mathrm{FF}}+1)^{2}-1]/2$ and $(J/F)\sqrt{D_{f}/{D_{v}}}\equiv\widetilde{\mathcal{T}}_{\mathrm{FB}}=\sqrt{(2\mathcal{T}_{\mathrm{FB}}+1)^{2}-1}$, while the last collective variable, $F/H\equiv\rho$, denotes the relative speed of transmission of the two information. At weak signal strengths, we have the further simplification $\widetilde{\mathcal{T}}_{\mathrm{FF}}=2\mathcal{T}_{\mathrm{FF}}$.
Substituting the collective variables for information gives the BEST in shallow gradients,
\begin{align}
\frac{\mathcal{P}}{\mathcal{P}_{0}}=\frac{\mathcal{T}_{\mathrm{FF}}}{\frac{1}{2}\left[ \sqrt{(2\mathcal{T}_{\mathrm{FB}}+1)^{2}-1}+\frac{1}{\sqrt{(2\mathcal{T}_{\mathrm{FB}}+1)^{2}-1}}\right] }\label{eqs:bestshallow},
\end{align}
and the BEST for all gradients,
\begin{align}
\frac{\mathcal{P}}{\mathcal{P}_{0}}=\frac{\widetilde{\mathcal{T}}_{\mathrm{FF}}\widetilde{\mathcal{T}}_{\mathrm{FB}}\rho[1+\rho-\widetilde{\mathcal{T}}_{\mathrm{FF}}\widetilde{\mathcal{T}}_{\mathrm{FB}}]}{\rho(1+\rho)(1+\widetilde{\mathcal{T}}_{\mathrm{FB}}^{2})+\widetilde{\mathcal{T}}_{\mathrm{FF}}\widetilde{\mathcal{T}}_{\mathrm{FB}}}\label{eqs:beststeep}.
\end{align}

\subsection*{Information-theoretic bounds on performance}
The BEST equation quantifies how the information-theoretic collective variables act as resources that enable and limit navigation. The effect of limiting each different resource is best demonstrated by deriving a series of tight bounds on the performance starting from BEST.

\subsubsection*{Feedforward information strength} In shallow gradients, the denominator of the BEST, from Eq. \ref{eqs:bestshallow}, is of the form $p+1/p$ for $p>0$. Applying the AM-GM inequality $p+1/p\geq 2$, we get the first bound,
\begin{align}
\frac{\mathcal{P}}{\mathcal{P}_{0}}\leq \mathcal{T}_{\mathrm{FF}}.
\end{align}
This bound is saturated at the optimal feedback information $\mathcal{T}_{\mathrm{FB}}^{*}$ that satisfies $\sqrt{(2\mathcal{T}^{*}_{\mathrm{FB}}+1)^{2}-1}=1$, i.e., $\mathcal{T}_{\mathrm{FB}}^{*}=(\sqrt{2}-1)/2~\mathrm{nats}$.

\subsubsection*{Feedback information strength} 
\newcommand{\tfb}{\mathcal{T}_{\mathrm{FB}}}
The denominator of the BEST, from Eq. \ref{eqs:bestshallow}, also satisfies,
\begin{align}
\frac{1}{2}\left[ \sqrt{(2\mathcal{T}_{\mathrm{FB}}+1)^{2}-1}+\frac{1}{\sqrt{(2\mathcal{T}_{\mathrm{FB}}+1)^{2}-1}}\right] &= \frac{1}{2}\left[ 2\sqrt{\mathcal{T}_{\mathrm{FB}}}\sqrt{\mathcal{T}_{\mathrm{FB}}+1}+\frac{1}{2\sqrt{\mathcal{T}_{\mathrm{FB}}}\sqrt{\mathcal{T}_{\mathrm{FB}}+1}}\right] \\
&=\frac{4\tfb(\tfb+1)+1}{4\sqrt{\tfb}\sqrt{\tfb+1}}\\
&=\frac{1}{4\sqrt{\tfb}}\cdot\sqrt{\frac{1+8\tfb+24\tfb^{2}+32\tfb^{3}+16\tfb^{4}}{1+\tfb}}\\
&\geq\frac{1}{4\sqrt{\tfb}}\cdot\sqrt{\frac{1+\tfb}{1+\tfb}}=\frac{1}{4\sqrt{\tfb}}.
\end{align}
This gives the second bound,
\begin{align}
\frac{\mathcal{P}}{\mathcal{P}_{0}}\leq 4\mathcal{T}_{\mathrm{FF}}\sqrt{\tfb},
\end{align}
which is saturated in the limit $\tfb\to 0$.

The denominator of Eq. \ref{eqs:bestshallow} also satisfies,
\begin{align}
\frac{1}{2}\left[ \sqrt{(2\mathcal{T}_{\mathrm{FB}}+1)^{2}-1}+\frac{1}{\sqrt{(2\mathcal{T}_{\mathrm{FB}}+1)^{2}-1}}\right] &\geq \frac{1}{2}\left[ \sqrt{(2\mathcal{T}_{\mathrm{FB}}+1)^{2}-1}\right] \geq \frac{1}{2}\left[ \sqrt{4\mathcal{T}_{\mathrm{FB}}^{2}}\right] =\tfb.
\end{align}
This gives the third bound,
\begin{align}
\frac{\mathcal{P}}{\mathcal{P}_{0}}\leq \frac{\mathcal{T}_{\mathrm{FF}}}{\tfb},
\end{align}
which becomes tight in the asymptotic limit of high $\tfb$.

The three above bounds are tight in three regimes: at low, intermediate and high feedback information. They can be summarized into a single expression,
\begin{align}
\frac{\mathcal{P}}{\mathcal{P}_{0}}\leq {\rm MIN} \{4 \mathcal{T}_{\mathrm{FF}} \sqrt{\mathcal{T}_{\mathrm{FB}}},\mathcal{T}_{\mathrm{FF}}, \frac{\mathcal{T}_{\mathrm{FF}}}{\mathcal{T}_{\mathrm{FB}}}\}.
\end{align}

\subsubsection*{Relative timescales}
In steep gradients, the relative speed of sensing over actuation, $\rho=F/H$, is another fundamental resource for navigation besides the strengths of information. This can be seen from Eq. \ref{eqs:beststeep}, which satisfies
\begin{align}
\frac{\mathcal{P}}{\mathcal{P}_{0}}\leq\frac{\widetilde{\mathcal{T}}_{\mathrm{FF}}\widetilde{\mathcal{T}}_{\mathrm{FB}}\rho[1+\rho]}{\rho(1+\rho)(1+\widetilde{\mathcal{T}}_{\mathrm{FB}}^{2})}=\frac{\widetilde{\mathcal{T}}_{\mathrm{FF}}}{\left( \widetilde{\mathcal{T}}_{\mathrm{FB}}+\frac{1}{\widetilde{\mathcal{T}}_{\mathrm{FB}}}\right)},
\end{align}
which is the performance in shallow gradients for the same strengths of bidirectional information. 

Additionally, Eq. \ref{eqs:beststeep} is a monotonically increasing function in $\rho$ for $\rho\geq\widetilde{\mathcal{T}}_{\mathrm{FF}}\widetilde{\mathcal{T}}_{\mathrm{FB}}-1$, which is the regime for stability. This can be seen by taking the derivative of Eq. \ref{eqs:beststeep} with respect to $\rho$. This gives, after simplification,
\begin{align}
\frac{d}{d\rho}\frac{\mathcal{P}}{\mathcal{P}_{0}}=\frac{\widetilde{\mathcal{T}}_{\mathrm{FF}}^{2}\widetilde{\mathcal{T}}_{\mathrm{FB}}^{2}[\rho(1+\rho)+\widetilde{\mathcal{T}}_{\mathrm{FB}}^{2}\rho^{2}+(1+\rho-\widetilde{\mathcal{T}}_{\mathrm{FF}}\widetilde{\mathcal{T}}_{\mathrm{FB}})]}
{[\rho(1+\rho)(1+\widetilde{\mathcal{T}}_{\mathrm{FB}}^{2})+\widetilde{\mathcal{T}}_{\mathrm{FF}}\widetilde{\mathcal{T}}_{\mathrm{FB}}]^{2}}\geq 0,
\end{align}
since $1+\rho\geq\widetilde{\mathcal{T}}_{\mathrm{FF}}\widetilde{\mathcal{T}}_{\mathrm{FB}}$. 

Therefore, when the strengths of the feedforward and feedback information flows are kept fixed, the performance increases monotonically with $\rho$ to saturate at the shallow gradient performance in the asymptotic limit of large $\rho$.

\newpage

\section*{Temporal sensing: nonlinear perturbation theory for information and performance}

We show here the full derivation of the analytical expressions for the information and the navigation performance in the minimal model for temporal sensing,
\begin{align}
\dot{f}&=-Ff+Ggv+\sqrt{2D_{f}}\psi\label{eqs:tempmodel1},\\
\dot{v}&=-Hv+Jfv+\sqrt{2D_{v}}\eta\label{eqs:tempmodel2}.
\end{align}
In this model, the bidirectional information transmission is defined through the transfer entropy rates,
\begin{align}
\dot{T}_{v\to f}&=\lim_{\delta t\to0}\frac{1}{\delta t}\left\langle\log\frac{P[f(t+\delta t)|\mathbf{f},\mathbf{v}]}{P[f(t+\delta t)|\mathbf{f}]}\right\rangle , \mathcal{T}_{\mathrm{FF}}=\dot{T}_{v\to f}/H,\label{eqs:tvf}\\
\dot{T}_{f\to v}&=\lim_{\delta t\to0}\frac{1}{\delta t}\left\langle\log\frac{P[v(t+\delta t)|\mathbf{f},\mathbf{v}]}{P[v(t+\delta t)|\mathbf{v}]}\right\rangle , \mathcal{T}_{\mathrm{FB}}=\dot{T}_{f\to v}/F,\label{eqs:tfv}
\end{align}
and the performance is defined as the drift speed,
\begin{align}
\mathcal{P}=\langle v\rangle\;,\;\mathcal{P}_{0}=\sigma_{v0},\label{eqs:tempperf}
\end{align}
where the angular brackets $\langle\cdot\rangle$ denote a steady-state average, and $\sigma_{v0}$ is the standard deviation of $v$ at steady-state when $J=0$.

Due to the nonlinear term in Eq. \ref{eqs:tempmodel2}, $Jfv$, the steady-state joint distribution of the trajectories resulting from the dynamics is not Gaussian. Therefore, the information and performance cannot be derived analytically in closed form. 

Our central innovation is the development of a controlled perturbation theory in the regime of weak nonlinearity. This theory has enabled the derivation of analytical expressions of information and performance that are valid up to second order in nonlinearity, i.e., up to $\mathcal{O}(J^{3})$. We show this derivation below.

\subsection*{Simplification for Gaussian noise}
The first step in deriving the transfer entropy rates in Eqs. \ref{eqs:tvf} and \ref{eqs:tfv} is to simplify them using the Langevin equations \ref{eqs:tempmodel1} and \ref{eqs:tempmodel2}. The white noise sources $\psi$ and $\eta$ are Gaussian distributed. This implies that for an infinitesimal time increment $\delta t$, the increments in  $f$ and $v$ have the Gaussian distributions,
\begin{align}
P[f(t+\delta t)|\mathbf{f},\mathbf{v}]&=\frac{1}{\sqrt{4\pi\delta t D_{f}}}\exp \left[ -\frac{[f(t+\delta t)-f(t)+Ff(t)\delta t-Ggv(t)\delta t]^{2}}{4D_{f}\delta t}\right] \equiv\frac{1}{\sqrt{4\pi\delta t D_{f}}}\exp \left[ -\frac{\delta f^{2}}{4D_{f}\delta t}\right] ,\label{eqs:transformedf1}\\
P[v(t+\delta t)|\mathbf{f},\mathbf{v}]&=\frac{1}{\sqrt{4\pi\delta t D_{v}}}\exp \left[ -\frac{[v(t+\delta t)-v(t)+Hv(t)\delta t-Jf(t)v(t)\delta t]^{2}}{4D_{v}\delta t}\right] \equiv\frac{1}{\sqrt{4\pi\delta t D_{v}}}\exp \left[ -\frac{\delta v^{2}}{4D_{v}\delta t}\right] \label{eqs:transformedv1}.
\end{align}
When only $\mathbf{f}$ is known instead of both $\mathbf{f}$ and $\mathbf{v}$, the increments of $f(t)$ have the distribution,
\begin{align}
P[f(t+\delta t)|\mathbf{f}]&=\int D[\mathbf{v}'] P(f(t+\delta t)|\mathbf{f},\mathbf{v}') P(\mathbf{v}'|\mathbf{f})\\
&=\frac{1}{\sqrt{4\pi\delta t D_{f}}}\int dv'(t)\;\exp \left[ -\frac{[f(t+\delta t)-f(t)+Ff(t)\delta t-Ggv'(t)\delta t]^{2}}{4D_{f}\delta t}\right]
P[v'(t)|\mathbf{f}]\\
&=\frac{1}{\sqrt{4\pi\delta t D_{f}}}\int dv'(t)\;\exp \left[ -\frac{[\delta f-Gg(v'-v)\delta t]^{2}}{4D_{f}\delta t}\right]
P[v'(t)|\mathbf{f}] \label{eqs:transformedf2}.
\end{align}

Dividing Eq. \ref{eqs:transformedf2} by Eq. \ref{eqs:transformedf1} gives
\begin{align}
\frac{P[f(t+\delta t)|\mathbf{f}]}{P[f(t+\delta t)|\mathbf{f},\mathbf{v}]}&=\int dv'(t)\;\exp\left[ -\frac{G^{2}g^{2}(v'-v)^{2}\delta t}{4D_{f}}+\frac{Gg(v'-v)\delta f}{2D_{f}}\right] P[v'(t)|\mathbf{f}].
\end{align}
Since $\delta t$ is infinitesimal, and $\delta f\sim\sqrt{\delta t}$, we can Taylor expand the above equation to obtain, up to linear order in $\delta t$,
\begin{align}
\frac{P[f(t+\delta t)|\mathbf{f}]}{P[f(t+\delta t)|\mathbf{f},\mathbf{v}]}&=\int dv'(t)\;\left[ 1-\frac{G^{2}g^{2}(v'-v)^{2}\delta t}{4D_{f}}+\frac{Gg(v'-v)\delta f}{2D_{f}}+\frac{G^{2}g^{2}(v'-v)^{2}\delta f^{2}}{8D_{f}^{2}}\right] P[v'(t)|\mathbf{f}]\\
&=1-\frac{G^{2}g^{2}\langle (v'-v)^{2}\rangle_{v'|\mathbf{f}}\delta t}{4D_{f}}+\frac{Gg\langle v'-v\rangle_{v'|\mathbf{f}}\delta f}{2D_{f}}+\frac{G^{2}g^{2}\langle (v'-v)^{2}\rangle_{v'|\mathbf{f}}\delta f^{2}}{8D_{f}^{2}},
\end{align}
where the conditional average over the distribution $P[v'(t)|\mathbf{f}]$ is denoted by the angular brackets $\langle\cdot\rangle_{v'|\mathbf{f}}$.
We take a logarithm and further Taylor expand using $\log(1+X)=X-X^{2}/2+\dots$. Retaining terms up to linear order in $\delta t$, i.e., up to $\delta f^{2}$, we get
\begin{align}
\log\frac{P[f(t+\delta t)|\mathbf{f}]}{P[f(t+\delta t)|\mathbf{f},\mathbf{v}]}=-\frac{G^{2}g^{2}\langle (v'-v)^{2}\rangle_{v'|\mathbf{f}}\delta t}{4D_{f}}+\frac{Gg\langle v'-v\rangle_{v'|\mathbf{f}}\delta f}{2D_{f}}+\frac{G^{2}g^{2}\langle (v'-v)^{2}\rangle_{v'|\mathbf{f}}\delta f^{2}}{8D_{f}^{2}}-\frac{G^{2}g^{2}\langle v'-v\rangle_{v'|\mathbf{f}}^{2}\delta f^{2}}{8D_{f}^{2}}.
\end{align}
Finally, we need to average over the conditioned trajectories, $\mathbf{f}$ and $\mathbf{v}$, in order to find the transfer entropy expression in Eq. \ref{eqs:tvf}. We exploit
$\langle\delta f\rangle=0$, $\langle \delta f^{2}\rangle=2D_{f}\delta t$, and that $\delta f$ is independent from both $\mathbf{f}$ and $\mathbf{v}$, to get
\begin{align}
\left\langle\log\frac{P[f(t+\delta t)|\mathbf{f}]}{P[f(t+\delta t)|\mathbf{f},\mathbf{v}]}\right\rangle&=-\frac{G^{2}g^{2}\left\langle \langle (v'-v)^{2}\rangle_{v'|\mathbf{f}}\right\rangle \delta t}{4D_{f}}+0+\frac{G^{2}g^{2}\left\langle \langle (v'-v)^{2}\rangle_{v'|\mathbf{f}}\right\rangle \delta t}{4D_{f}}-\frac{G^{2}g^{2}\left\langle \langle v'-v\rangle_{v'|\mathbf{f}}^{2}\right\rangle \delta t}{4D_{f}}\\
&=-\frac{G^{2}g^{2}\left\langle \langle v'-v\rangle_{v'|\mathbf{f}}^{2}\right\rangle \delta t}{4D_{f}}\\
&=-\frac{G^{2}g^{2}\left\langle (\langle v'\rangle_{v'|\mathbf{f}}-v)^{2}\right\rangle \delta t}{4D_{f}}=-\frac{G^{2}g^{2}\delta t}{4D_{f}}\left\langle \mathrm{Var}\left( P(v(t)|\mathbf{f}_{[-\infty,t]})\right)\right\rangle ,
\end{align}
where, in the penultimate step, we have used the independence of $v(t)$ from the conditional average over the dummy variable $v'(t)$. The last step highlights that the outermost averaging operation in $\langle(\langle v'\rangle_{v'|\mathbf{f}}-v)^{2}\rangle$ can be split into two steps, $\langle(\langle v'\rangle_{v'|\mathbf{f}}-v)^{2}\rangle=\langle\langle(\langle v'\rangle_{v'|\mathbf{f}}-v)^{2}\rangle_{\mathbf{v}}\rangle_{\mathbf{f}}$, giving the conditional variance of $v$, $\langle(\langle v'\rangle_{v'|\mathbf{f}}-v)^{2}\rangle_{\mathbf{v}}=\mathrm{Var}\left( P(v(t)|\mathbf{f}_{[-\infty,t]})\right)$, averaged over the conditioned trajectory $\mathbf{f}$.

Substituting in Eq. \ref{eqs:tvf}, the feedforward transfer entropy rate is given by
\begin{align}
\dot{\mathcal{T}}_{v\to f}=\frac{G^{2}g^{2}}{4D_{f}}\left\langle \mathrm{Var}\left( P(v(t)|\mathbf{f}_{[-\infty,t]})\right)\right\rangle \equiv\frac{G^{2}g^{2}}{4D_{f}}\left\langle P^{v|f}\right\rangle\label{eqs:girsanov1},
\end{align}
where $P^{v|f}$ is defined to be the conditional variance.

An analogous derivation gives the feedback transfer entropy rate as
\begin{align}
\dot{\mathcal{T}}_{f\to v}&=\frac{J^{2}}{4D_{v}}\left\langle v^{2}\mathrm{Var}\left( P(f(t)|\mathbf{v}_{[-\infty,t]})\right)\right\rangle\equiv\frac{J^{2}}{4D_{v}}\left\langle v^{2}P^{f|v}\right\rangle\label{eqs:girsanov2},
\end{align}
where $P^{f|v}$ is defined to be, again, the conditional variance. 

The above expressions for the transfer entropy rates are exact. Next, we discuss how to calculate the conditional variances from the equations of motion.

\subsection*{Stochastic Riccati equations}
Since the nonlinear coupling term in the equations of motion, $Jfv$, has a bilinear structure, the  process $\mathbf{f}$ conditioned on the trajectory $\mathbf{v}$ and the process $\mathbf{v}$ conditioned on the trajectory $\mathbf{f}$ are both linear Ornstein-Uhlenbeck processes. The former has a time-dependent center given by $Ggv(t)/F$ and a constant spring-constant $F$, while the latter has a constant center at 0 and a time-dependent spring-constant given by $H-Jf(t)$. Therefore, the conditional probability distributions $P(f(t)|\mathbf{v}_{[-\infty,t]})$ and $P(v(t)|\mathbf{f}_{[-\infty,t]})$ are Gaussian, even though the joint distribution $P(\mathbf{f},\mathbf{v})$ is not Gaussian. 

The Gaussian conditional distributions satisfy the \textit{stochastic} Riccati equations \cite{jazwinski2007stochastic},
\begin{align}
\dot{P}^{v|f}&=2[-H+Jf(t)]P^{v|f}+2D_{v}-\frac{G^{2}g^{2}}{2D_{f}}(P^{v|f})^{2}\label{eqs:sr1},\\
\dot{P}^{f|v}&=-2FP^{f|v}+2D_{f}-\frac{J^{2}v(t)^{2}}{2D_{v}}(P^{f|v})^{2}\label{eqs:sr2}.
\end{align}
If we could solve these equations for $P^{v|f}$ and $P^{f|v}$, we could use those expressions to get the transfer entropy rates through Eqs. \ref{eqs:girsanov1} and \ref{eqs:girsanov2}.

\subsection*{Exact solution in $J=0$ limit}Note that the Riccati equations become exactly solvable when the nonlinear term is zero, i.e., when $J=0$. In that limit, Eqs. \ref{eqs:sr1} and \ref{eqs:sr2} become
\begin{align}
\dot{P}^{v|f}_{0}&=-2HP^{v|f}_{0}+2D_{v}-\frac{G^{2}g^{2}}{2D_{f}}(P^{v|f}_{0})^{2}\label{eqs:linsr1},\\
\dot{P}^{f|v}_{0}&=-2FP^{f|v}_{0}+2D_{f}\label{eqs:linsr2},
\end{align}
where the conditional variances in the $J=0$ limit are denoted as $P^{v|f}_0$ and $P^{f|v}_0$. These equations do not depend on the respective conditioned observable; all the coefficients are constants. Therefore, for any given conditioned trajectories $\mathbf{f}$ and $\mathbf{v}$, they evolve towards the same fixed points. These fixed points are given by putting the LHS in the above equations to zero, which yields the positive solutions as,
\begin{align}
P_{0}^{v|f*}&=\frac{2D_{f}}{G^{2}g^{2}}\left[ \sqrt{H^{2}+\frac{G^{2}g^{2}D_{v}}{D_{f}}}-H\right] ,\label{eqs:pvf0}\\
P_{0}^{f|v*}&= \frac{D_{f}}{F} \label{eqs:pfv0}.
\end{align}
Since $P^{v|f}_{0}$ and $P^{f|v}_{0}$ in steady-state do not depend on the specific realization of the respective conditioned trajectory, the steady-state average is the same as the fixed points. This implies that the feedforward transfer entropy rate in this limit is
\begin{align}
\dot{\mathcal{T}}_{v\to f}^{(0)}=\frac{G^{2}g^{2}}{4D_{f}}P_{0}^{v|f*}=\frac{1}{2}\left[ \sqrt{H^{2}+\frac{G^{2}g^{2}D_{v}}{D_{f}}}-H\right] .
\end{align}
Additionally, since $J=0$, the feedback transfer entropy rate is
\begin{align}
\dot{\mathcal{T}}_{f\to v}^{(0)}=0.
\end{align}

In this limit, the performance is also trivially solvable. When $J=0$, the equation of motion of $v$ has no coupling with $f$. Taking a steady-state average, and using $\langle\dot{v}\rangle=0$ due to stationarity, gives the performance as
$\mathcal{P}=\langle v_{0}\rangle=0$. The unit of the performance, $\mathcal{P}_{0}=\sigma_{v0}$, is an element of the steady-state covariance matrix $\mathbf{C}_{0}$, which solves the Lyapunov equation
$\mathbf{AC_{0}}+\mathbf{C_{0}A}^{T}=2\mathbf{D}$, where $\mathbf{A}=\begin{pmatrix}F&-Gg\\0&H\end{pmatrix}$ is the spring-constant matrix and $\mathbf{D}=\begin{pmatrix}D_{f}&0\\0&D_{v}\end{pmatrix}$ is the diffusion-constant matrix. The solution is given by,
\begin{align}
\mathbf{C}_{0}=\begin{pmatrix}\frac{D_{f}}{F}+\frac{G^{2}g^{2}D_{v}}{FH(F+H)}&\frac{GgD_{v}}{H(F+H)}\\\frac{GgD_{v}}{H(F+H)}&\frac{D_{v}}{H}\end{pmatrix}\label{eqs:tempvarcovar}.
\end{align}
The second diagonal element gives $\sigma_{v0}=\sqrt{D_{v}/H}$.

\subsection*{Perturbative solution for $J>0$}
For a weak nonlinear coupling, we derive a perturbative solution for the Riccati equations around $J=0$. We construct a perturbation expansion in $J$ for all the stochastic variables in the Riccati equations,
\begin{align}
P^{v|f}(t)&=P_{0}^{v|f}(t)+JP_{1}^{v|f}(t)+J^{2}P_{2}^{v|f}(t)+\dots,\label{eqs:pvfpert}\\
P^{f|v}(t)&=P_{0}^{f|v}(t)+JP_{1}^{f|v}(t)+J^{2}P_{2}^{f|v}(t)+\dots,\label{eqs:pfvpert}\\
f(t)&=f_{0}(t)+Jf_{1}(t)+J^{2}f_{2}(t)+\dots,\label{eqs:fpert}\\
v(t)&=v_{0}(t)+Jv_{1}(t)+J^{2}v_{2}(t)+\dots\label{eqs:vpert}.
\end{align}
Since we perform this expansion for not only the conditional variances but also the stochastic trajectories themselves, the steady-state averages $\langle\cdot\rangle$ in Eqs. \ref{eqs:girsanov1} and \ref{eqs:girsanov2} should be interpreted henceforth as averages over the Gaussian noise measure, i.e., independent of $J$.

In order to solve for the unknown terms at each order of $J$ in the above expansions, we substitute the perturbation expansions into the Riccati equations, Eqs. \ref{eqs:sr1} and \ref{eqs:sr2}, and the equations of motion, Eqs. \ref{eqs:tempmodel1} and \ref{eqs:tempmodel2}, and solve at each order in $J$. This gives us analytical expressions for $P^{v|f}$, $P^{f|v}$ and $v$ accurate up to a given order in $J$. We substitute them in Eqs. \ref{eqs:girsanov1}, \ref{eqs:girsanov2} and \ref{eqs:tempperf}, respectively, to obtain the transfer entropy rates and the performance. 
The solutions for each case are described below.

\subsubsection*{Performance}
The performance $\mathcal{P}=\langle v\rangle$ can be derived by substituting the perturbation expansion for $v$, Eq. \ref{eqs:vpert}, into the RHS of the equation of motion for $v$, Eq. \ref{eqs:tempmodel2}. This gives,
\begin{align}
\dot{v}=-H(v_{0}+Jv_{1}+J^{2}v_{2})+J(f_{0}+Jf_{1}+J^{2}f_{2})(v_{0}+Jv_{1}+J^{2}v_{2})+\sqrt{2D_{v}}\eta.
\end{align}
At steady-state, the average of $v$ becomes stationary, i.e., $\langle\dot{v}\rangle=0$.
Therefore, taking the steady-state average of the above equation and matching the terms for each order in $J$, up to linear order, gives,
\begin{align}
\langle v_{0}\rangle&=0,\\
\langle v_{1}\rangle&=\frac{1}{H}\langle f_{0}v_{0}\rangle=\frac{GgD_{v}}{H^{2}(F+H)}\label{eqs:perflin}.
\end{align}
where, in the last equation, we have used the exact solution for $J=0$ from Eq. \ref{eqs:tempvarcovar}. 
The second-order contribution vanishes due to a parity symmetry in the system. The equations of motion, Eqs. \ref{eqs:tempmodel1} and \ref{eqs:tempmodel2}, are invariant under the transformation $(f,v,J)\to(-f,-v,-J)\equiv (\tilde{f},\tilde{v},\tilde{J})$. Therefore, $\langle f\rangle_{J}=\langle\tilde{f}\rangle_{\tilde{J}}=-\langle f\rangle_{-J}$, and $\langle v\rangle_{J}=\langle\tilde{v}\rangle_{\tilde{J}}=-\langle v\rangle_{-J}$. Indeed, $\langle f\rangle$ and $\langle v\rangle$ are both odd in $J$. Consequently, the averages of the even terms in Eqs. \ref{eqs:fpert} and \ref{eqs:vpert} are zero, i.e, $\langle f_{0}\rangle=\langle f_{2}\rangle=\langle v_{0}\rangle=\langle v_{2}\rangle=0$. Therefore, the performance up to second order in $J$ is
\begin{align}
\mathcal{P}=J \langle v_1 \rangle +\mathcal{O}(J^{3})= \frac{JGgD_{v}}{H^{2}(F+H)}+\mathcal{O}(J^{3}).
\end{align}

\subsubsection*{Feedforward transfer entropy rate}
The feedforward transfer entropy rate depends on $\langle P^{v|f}\rangle$ as seen in Eq. \ref{eqs:girsanov1}. To obtain this quantity, first, we substitute the perturbation expansion for $P^{v|f}$, Eq. \ref{eqs:pvfpert}, and the perturbation expansion for $f(t)$, Eq. \ref{eqs:fpert}, in the corresponding Riccati equation, Eq. \ref{eqs:sr1}, to get, at each order in $J$,
\begin{align}
\dot{P}_{0}^{v|f}&=-2H P_{0}^{v|f}+2D_{v}-\frac{G^{2}g^{2}}{2D_{f}}(P_{0}^{v|f})^{2}\label{eqs:pvfa},\\
\dot{P}_{1}^{v|f}&=-2H P_{1}^{v|f}+2f_{0}P_{0}^{v|f}-\frac{G^{2}g^{2}}{D_{f}}P_{0}^{v|f}P_{1}^{v|f}\label{eqs:pvfb},\\
\dot{P}_{2}^{v|f}&=-2H P_{2}^{v|f}+2f_{1}P_{0}^{v|f}+2f_{0}P_{1}^{v|f}-\frac{G^{2}g^{2}}{2D_{f}}(P_{1}^{v|f})^{2}-\frac{G^{2}g^{2}}{D_{f}}P_{0}^{v|f}P_{2}^{v|f}\label{eqs:pvfc}.
\end{align}
We will now solve them at each order. Note that while the transfer entropy depends on only the steady-state average $\langle P^{v|f}\rangle$, the above equations at each order depend on the full time-dependent solution at the previous order. Therefore, the first two equations need to be solved with the time-dependence, while the last equation needs to be solved for only the steady-state average. 

The first equation, Eq. \ref{eqs:pvfa}, is the $J=0$ case that was solved in the previous section. $P_{0}^{v|f}$ relaxes towards the fixed point $P_{0}^{v|f*}$ given in Eq. \ref{eqs:pvf0}, independent of the stochastic dynamics of the conditioned trajectory $\mathbf{f}$. Therefore, in steady-state, $P_{0}^{v|f}(t)$ will be already at the fixed point without any time-dependence. We will henceforth substitute $P_{0}^{v|f}(t)$ by $P_{0}^{v|f*}$.

Using this, the second equation, Eq. \ref{eqs:pvfb}, has the linear form,
\begin{align}
\dot{P}_{1}^{v|f}=-\lambda P_{1}^{v|f}+2f_{0}(t) P_{0}^{v|f*},\label{eqs:p1lin}
\end{align}
where $\lambda\equiv 2H+G^{2}g^{2}P_{0}^{v|f*}/D_{f}$ is the auto-regression rate for $P_{1}^{v|f}(t)$, and $2f_{0}(t)P_{0}^{v|f*}$ is a time-dependent source term. The solution at steady-state is
\begin{align}
P_{1}^{v|f}(t)&=2P_{0}^{v|f*}\int_{-\infty}^{t}ds\;e^{-\lambda (t-s)}f_{0}(s)\label{eqs:p1sol}\\
\implies \langle P_{1}^{v|f}(t)\rangle&=2P_{0}^{v|f*}\int_{-\infty}^{t}ds\;e^{-\lambda (t-s)}\langle f_{0}(s)\rangle.
\end{align}
This vanishes due to the parity symmetry discussed before. Therefore,
\begin{align}
\langle P_{1}^{v|f}(t)\rangle=0\label{eqs:pvf1}.
\end{align}

The third of the perturbed Riccati equations, Eq. \ref{eqs:pvfc}, needs to be solved for only the steady-state average, $\langle P^{v|f}_{2}\rangle$, instead of the full time-dependent $P^{v|f}_{2}(t)$. By taking the steady-state average of Eq. \ref{eqs:pvfc}, and using $\langle\dot{P}^{v|f}_{2}\rangle=0$ in the stationary state, we get
\begin{align}
0=-\lambda\langle P_{2}^{v|f}\rangle+2P_{0}^{v|f*}\langle f_{1}\rangle+2\langle f_{0}P_{1}^{v|f}\rangle-\frac{G^{2}g^{2}}{2D_{f}}\langle (P_{1}^{v|f})^{2}\rangle\label{eqs:p2av},
\end{align}
where we have again substituted $2H+G^{2}g^{2}P_{0}^{v|f*}/D_{f}=\lambda$. The last two terms can be shown to be proportional to each other, by multiplying Eq. \ref{eqs:p1lin} by $P_{1}^{v|f}$, which gives,
\begin{align}
&P_{1}^{v|f}\dot{P}_{1}^{v|f}=-\lambda (P_{1}^{v|f})^{2}+2P_{0}^{v|f*}f_{0}P_{1}^{v|f}\\
\implies&\frac{1}{2}\frac{d}{dt}(P_{1}^{v|f})^{2}=-\lambda (P_{1}^{v|f})^{2}+2P_{0}^{v|f*}f_{0}P_{1}^{v|f}\\
\implies&0=-\lambda\langle(P_{1}^{v|f})^{2}\rangle+2P_{0}^{v|f*}\langle f_{0}P_{1}^{v|f}\rangle,
\end{align}
where we have taken a steady-state average in the last step. Therefore, $\langle (P_{1}^{v|f})^{2}\rangle=2P_{0}^{v|f*}\langle f_{0}P_{1}^{v|f}\rangle/\lambda$. Substituting this in Eq. \ref{eqs:p2av} gives the second-order contribution as,
\begin{align}
\langle P_{2}^{v|f}\rangle=\frac{2P_{0}^{v|f*}}{\lambda}\langle f_{1}\rangle+\left[ \frac{2}{\lambda}-\frac{G^{2}g^{2}P_{0}^{v|f*}}{D_{f}\lambda^{2}}\right] \langle f_{0}P_{1}^{v|f}\rangle\label{eqs:p2twoterms}.
\end{align}

The first term in Eq. \ref{eqs:p2twoterms} can be calculated by substituting the perturbative expansion Eq. \ref{eqs:fpert} into the equation of motion \ref{eqs:tempmodel1} to get, up to second order in $J$,
\begin{align}
\dot{f}=-F(f_{0}+Jf_{1}+J^{2}f_{2})+Gg(v_{0}+Jv_{1}+J^{2}v_{2})+\sqrt{2D_{f}}\psi.
\end{align}
Taking the steady-state average of the above equation, using $\langle\dot{f}\rangle=0$, and matching the linear term in $J$ gives,
\begin{align}
\langle f_{1}\rangle=\frac{Gg}{F}\langle v_{1}\rangle=\frac{G^{2}g^{2}D_{v}}{H^{2}F(F+H)}\label{eqs:f1term},
\end{align}
where we have used $\langle v_{1}\rangle$ from Eq. \ref{eqs:perflin}.

The second term in Eq. \ref{eqs:p2twoterms} can be obtained from the time-dependent solution for $P_{1}^{v|f}$ in Eq. \ref{eqs:p1sol}. Multiplying Eq. \ref{eqs:p1sol} by $f_{0}(t)$ and taking a steady-state average gives,
\begin{align}
\langle f_{0}P_{1}^{v|f}\rangle&=2P_{0}^{v|f*}\int_{-\infty}^{t}ds\;e^{-\lambda(t-s)}\langle f_{0}(t)f_{0}(s)\rangle\label{eqs:f0p1},
\end{align}
which depends on the autocorrelation function for $f_{0}(t)$. Since the process $(f_{0},v_{0})$ is a linear process, the autocorrelation function is given by \cite{gardiner1985handbook},
\begin{align}
\langle f_{0}(t)f_{0}(s<t)\rangle=\left[ e^{-\mathbf{A}(t-s)}\mathbf{C}_{0}\right]_{ff}=e^{-F(t-s)}[\mathbf{C}_{0}]_{ff}+\frac{Gg[\mathbf{C}_{0}]_{fv}}{F-H}(e^{-H(t-s)}-e^{-F(t-s)}).
\end{align}
Substituting this in Eq. \ref{eqs:f0p1} gives,
\begin{align}
\langle f_{0}P_{1}^{v|f}\rangle&=2P_{0}^{v|f*}\int_{-\infty}^{t}ds\;e^{-\lambda(t-s)}\left[ e^{-F(t-s)}[\mathbf{C}_{0}]_{ff}+\frac{Gg[\mathbf{C}_{0}]_{fv}}{F-H}(e^{-H(t-s)}-e^{-F(t-s)})\right] \\
&=2P_{0}^{v|f*}\left[ \frac{[\mathbf{C}_{0}]_{ff}}{\lambda+F}+\frac{Gg[\mathbf{C}_{0}]_{fv}}{(\lambda+F)(\lambda+H)}\right] \label{eqs:f0p1term}.
\end{align}
Now we can substitute Eqs. \ref{eqs:f1term} and \ref{eqs:f0p1term} in Eq. \ref{eqs:p2twoterms} to get the full second-order perturbative correction,
\begin{align}
\langle P_{2}^{v|f}\rangle&=\frac{2P_{0}^{v|f*}}{\lambda}\cdot \frac{G^{2}g^{2}D_{v}}{H^{2}F(F+H)}+\left[ \frac{2}{\lambda}-\frac{G^{2}g^{2}P_{0}^{v|f*}}{D_{f}\lambda^{2}}\right] \cdot 2P_{0}^{v|f*}\left[ \frac{[\mathbf{C}_{0}]_{ff}}{\lambda+F}+\frac{Gg[\mathbf{C}_{0}]_{fv}}{(\lambda+F)(\lambda+H)}\right] 
\end{align}

This is the second-order contribution to $P^{v|f}$, with the zeroth and first-order contributions given by Eqs. \ref{eqs:pvf0} and \ref{eqs:pvf1}, respectively. We can substitute these in Eq. \ref{eqs:girsanov1}, use $\lambda=2H+G^{2}g^{2}P_{0}^{v|f*}/D_{f}$, $P_{0}^{v|f*}$ given by Eq. \ref{eqs:pvf0}, and $\mathbf{C}_{0}$ given by Eq. \ref{eqs:tempvarcovar}, to obtain, after simplification,
\begin{align}
\dot{\mathcal{T}}_{v\to f}&=\frac{1}{2}\left[ \sqrt{H^{2}+\frac{G^{2}g^{2}D_{v}}{D_{f}}}-H\right] +\frac{J^{2}D_{f}(S^{2}-H^{2})(H^{2}F-H^{2}S+2FS^{2}+4S^{3})}{2H^{2}FS(F+H)(F+2S)(H+2S)}+\mathcal{O}(J^{3}),
\end{align}
where $S$ is defined as $S\equiv \sqrt{H^{2}+G^{2}g^{2}D_{v}/D_{f}}$.

\subsubsection*{Feedback transfer entropy rate}
The expression for the feedback transfer entropy rate, Eq. \ref{eqs:girsanov2}, is proportional to $J^{2}\langle v^{2}P^{f|v}\rangle$. Therefore, the lowest order contribution to the feedback transfer entropy rate is
\begin{align}
\dot{\mathcal{T}}_{f\to v}=\frac{J^{2}}{4D_{v}}\langle v_{0}^{2}P_{0}^{f|v}\rangle +\mathcal{O}(J^{3}).
\end{align}
From the exact solution at $J=0$, we know that $P_{0}^{f|v}$ evolves towards the fixed point $P_{0}^{f|v*}$ independent from the stochastic dynamics. Therefore, the steady-state feedback transfer entropy rate is,
\begin{align}
\dot{\mathcal{T}}_{f\to v}=\frac{J^{2}}{4D_{v}}\langle v_{0}^{2}\rangle P_{0}^{f|v*} +\mathcal{O}(J^{3})=\frac{J^{2}}{4D_{v}}\cdot\frac{D_{v}}{H}\cdot\frac{D_{f}}{F} +\mathcal{O}(J^{3})=\frac{J^{2}D_{f}}{4HF}+\mathcal{O}(J^{3}),
\end{align}
where we have used $\langle v_{0}^{2}\rangle$ from the lower diagonal element of Eq. \ref{eqs:tempvarcovar}, and $P_{0}^{f|v*}$ from Eq. \ref{eqs:pfv0}.

\subsubsection*{Summary} We obtained analytical expressions for the transfer entropy rates and the performance up to second order in the nonlinear coupling $J$, by perturbatively solving the stochastic Riccati equations, Eqs. \ref{eqs:sr1} and \ref{eqs:sr2}, and the equations of motion, Eqs. \ref{eqs:tempmodel1} and \ref{eqs:tempmodel2}. 
This approach crucially relies on the stochastic Riccati equations being exact. This occurs when the nonlinear coupling has a bilinear functional form. Therefore, this approach can be applied to general bilinearly coupled dynamical systems to obtain controlled analytical approximations for the transfer entropy rates in the regime of weak nonlinearity.

\newpage
\section*{BEST for temporal sensing}
The BEST equation for temporal sensing relates the navigation performance to the strengths and timescales of bidirectional information transmission. We derive the BEST in the regime of shallow gradients and weak actuation strength, using the results of the perturbation theory.

The feedforward and feedback information and the performance, as derived in the previous section, are
\begin{align}
\mathcal{T}_{\mathrm{FF}}&=\dot{\mathcal{T}}_{v\to f}/H=\frac{1}{2}\left[ \sqrt{1+\frac{G^{2}g^{2}D_{v}}{H^{2}D_{f}}}-1\right] +\frac{J^{2}D_{f}(S^{2}-H^{2})(H^{2}F-H^{2}S+2FS^{2}+4S^{3})}{2H^{3}FS(F+H)(F+2S)(H+2S)}+\mathcal{O}(J^{3})\label{eqs:temporaltff}\\
&=\frac{G^{2}g^{2}D_{v}}{4H^{2}D_{f}}[1+\mathcal{O}(g^{2},J^{2})]\\
\mathcal{T}_{\mathrm{FB}}&=\dot{\mathcal{T}}_{f\to v}/F=\frac{J^{2}D_{f}}{4HF^{2}}+\mathcal{O}(J^{3}),\label{eqs:temporaltfb}\\
\mathcal{P}/\mathcal{P}_{0}&=\frac{JGg}{H(F+H)}\sqrt{\frac{D_{v}}{H}}+\mathcal{O}(J^{3})\label{eqs:temporalperf},
\end{align}
where $S=\sqrt{H^{2}+G^{2}g^{2}D_{v}/D_{f}}$.
The leading terms in each equation give,
\begin{align}
\mathcal{P}/\mathcal{P}_{0}=4\frac{F}{F+H}\left( \frac{Gg}{2H}\sqrt{\frac{D_{v}}{D_{f}}}\right)\left( \frac{J}{2F}\sqrt{\frac{D_{f}}{H}}\right) =4\frac{\rho}{1+\rho}\sqrt{\mathcal{T}_{\mathrm{FF}}}\sqrt{\mathcal{T}_{\mathrm{FB}}},
\end{align}
with $\rho=F/H$. This is the BEST for temporal sensing.

\newpage
\section*{Agent-based model for chemotaxis of \textit{Escherichia coli}}
The BEST equation in shallow gradient and weak actuation strength quantitatively predicts the simulated chemotactic performance of the bacterium \textit{E. coli} in an exponential concentration gradient. In this section, we specify how the parameters of our agent-based model relate to the parameters in a previous work by Long et al. \cite{long2017feedback}.

For convenience, we reproduce our model of the main text,
\begin{align}
\dot{f}&=-F(f-f_{0})+Ggv_{0}n\delta_{R}+\sqrt{2D_{f}}\psi,\label{eqs:Ecolif}\\
\lambda_{\mathrm{R}\to\mathrm{T}}(t)&=\omega_{_{R}}\mathrm{max}\{ [1-J(f-f_{0})],0\},\\
\lambda_{\mathrm{T}\to\mathrm{R}}(t)&=\omega_{T},\\
\dot{n}&=-n/(2D(\delta_R))+\sqrt{(1-n^{2})/(2D(\delta_R))}\;\eta.\label{eqs:Ecolin}
\end{align}
The sensory output $f$ is proportional to the free-energy difference between the active and inactive states of the sensory receptors. We follow Long et al. in setting this proportionality constant such that the probability to be in a run state, $\lambda_{\mathrm{T}\to \mathrm{R}}(J,f)/[\lambda_{\mathrm{R}\to \mathrm{T}}(J,f)+\lambda_{\mathrm{T}\to \mathrm{R}}(J,f)]$, is equal to $1/(1+e^{-f})$ when $J$ is at the typical value for wild-type cells, $J=J_{\mathrm{WT}}=0.32 \,{\rm s}^{-1}$
\cite{long2017feedback}. In these units, the sensory gain $G$ and the sensory noise magnitude $D_{f}$ are derived from those in Long et al. Additionally, $\omega_{R}$ and $\omega_{T}$ are derived by linearizing the switching propensities in \cite{long2017feedback} near the adapted value of $f$. Other parameters were taken without modification from \cite{long2017feedback}.

In order to apply our theory unambiguously, we also needed to specify the orientational diffusion constants of the cell such that the velocity autocorrelation function for $J=0$ has a single decay-rate. Fig. \ref{fig:sifig1}A shows the normalized velocity autocorrelation for a few values within the wild-type phenotypic distribution. We chose $D_{\rm run}=0.062\,{\rm s}^{-1}$, which exhibits a single decay-rate for the autocorrelation, and is close to the mean of the phenotypic distribution \cite{long2017feedback}.

\section*{Optimal feedback information in \textit{E. coli}}
The BEST for spatial sensing predicts an optimal amount of feedback information that maximizes navigation performance. Our perturbative theory for temporal sensing does not extend into this regime; we therefore instead used agent-based simulations to discover the existence of an optimal feedback information in \textit{E. coli}. In this section we show that the origin of this optimum in \textit{E. coli} is analogous to that in spatial sensing.

In spatial sensing, the optimal amount of feedback information originates from a trade-off between responding to the signal while avoiding  sensory noise. This can be seen from the expression for the optimal actuation strength, $J^{*}=F\sqrt{D_{v}/D_{f}}$. When the sensory noise $D_{f}$ decreases, $J^{*}$ increases. However, the optimal feedback information remains unchanged, $\mathcal{T}^{*}_{\mathrm{FB}}=(\sqrt{2}-1)/2$ nats, showing that information is a better collective variable for performance than the model parameters themselves.

\begin{figure}
\centering
\includegraphics[width=0.8\textwidth]{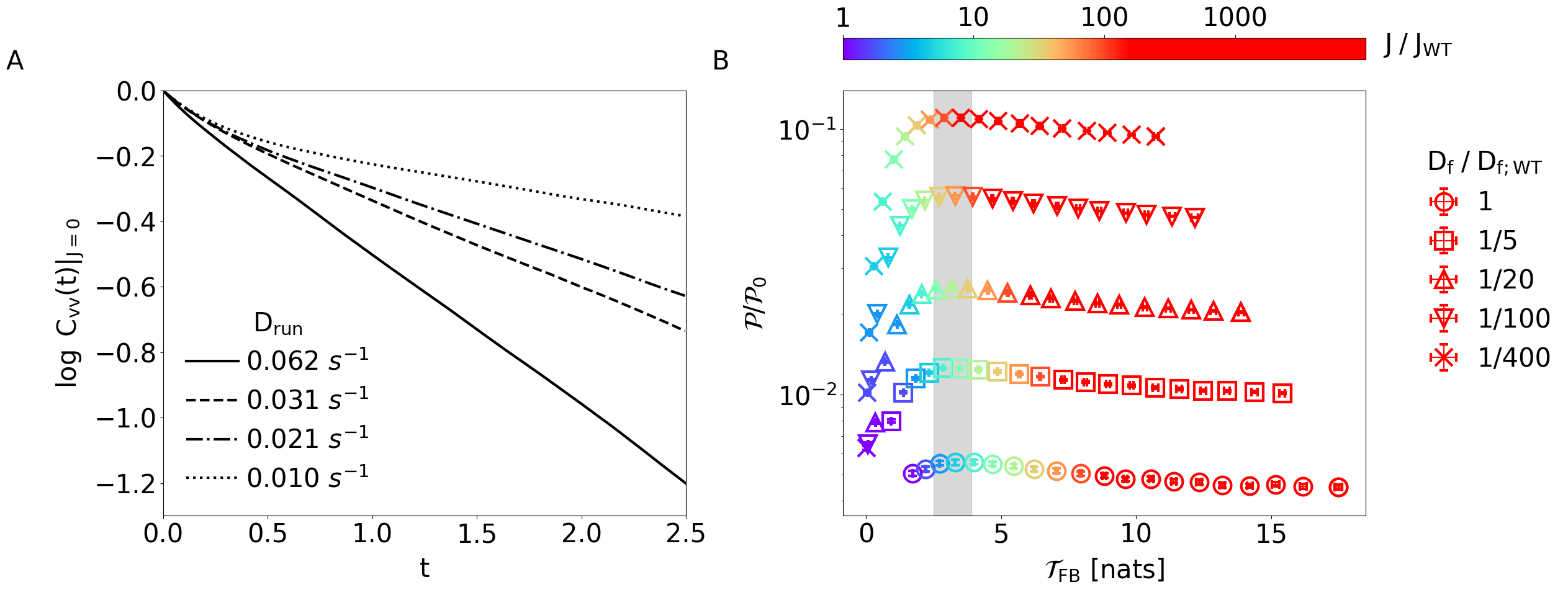}
\caption{(A) Normalized velocity autocorrelations at $J=0$ for different values of $D_{\mathrm{run}}$ within the wild-type phenotypic distribution of \textit{E. coli} \cite{long2017feedback}. In each case, $D_{\mathrm{tumble}}=37D_{\mathrm{run}}$. (B) Chemotactic performance of \textit{E. coli} as a function of feedback information. Feedback information is varied by changing the actuation strength $J$, indicated by color. The curves, marked by different symbols, correspond to simulations with different sensory noise magnitudes $D_f$ relative to the wild-type value, $D_{f,\mathrm{WT}}=0.096 \, \mathrm{s}^{-1}$. The shaded region indicates the regime of optimal feedback information.}
\label{fig:sifig1}
\end{figure}

\textit{E. coli} chemotaxis exhibits the same trade-off. Fig. \ref{fig:sifig1}B shows the navigation performance as a function of the feedback information for different values of the sensory noise magnitude $D_{f}$. For each value of $D_{f}$, we varied the actuation strength $J$ to tune the feedback information. When $D_f$ is reduced, the performance optimum occurs at a larger value of $J$: less sensory noise allows stronger actuation. However, when performance is plotted as a function of the feedback information, the optimum occurs at the same amount of feedback information for all values of $D_f$, even though the corresponding optimal value of $J$ changes. Thus, as in spatial sensing, the optimum is controlled by feedback information rather than by the microscopic actuation parameter itself. This supports the qualitative prediction of BEST.

\newpage

\section*{Single-step transfer entropy does not yield a BEST for spatial sensing}
The BEST equation shows that the bidirectional transfer entropy between the sensory input and output trajectories are the key collective variables that solely determine navigation performance. Instead, if the trajectories of sensory input and output are truncated to only their current instantaneous values for the calculation of the transfer entropy, it gives an approximation called the single-step transfer entropy \cite{chetrite2019information}. This approximation incurs an uncontrolled error. However, the single-step transfer entropy is still commonly used because it is easier to compute directly from experimental data.

For the spatial sensing model, we analytically show that the single-step approximation incurs a large, and sometimes diverging, error in the transfer entropy. Moreover, the navigation performance cannot be expressed as a function of solely the bidirectional single-step transfer entropies, i.e., as a BEST. Therefore, the bidirectional single-step transfer entropies are not the key collective variables that govern navigation.

\subsection*{Single-step transfer entropy between $x$ and $f$}
In the spatial sensing model, the single-step feedforward and feedback transfer entropy rates are defined, analogous to the multi-step transfer entropy in Eqs. \ref{eqs:spatialtvf} and \ref{eqs:spatialtfv}, as
\begin{align}
\dot{T}_{x\to f}^{(1)}&=\lim_{\delta t\to0+}\frac{1}{\delta t}\left\langle\log\frac{P[f(t+\delta t)|f(t),x(t)]}{P[f(t+\delta t)|f(t)]}\right\rangle ,\\
\dot{T}_{f\to x}^{(1)}&=\lim_{\delta t\to0+}\frac{1}{\delta t}\left\langle\log\frac{P[x(t+\delta t)|f(t),x(t)]}{P[x(t+\delta t)|x(t)]}\right\rangle .
\end{align}
The conditioning on the trajectories in Eqs. \ref{eqs:spatialtvf} and \ref{eqs:spatialtfv} have been replaced with the current values of $f(t)$ and $x(t)$. Note that single-step transfer entropy rates between $x$ and $f$ are not equal to those between $v$ and $f$, since the current values of $x$ and $v$ do not contain the same amount of information about $f$.

Due to the linearity of the spatial sensing system, the transfer entropy rates can be calculated through conditional variances,
\begin{align}
\dot{\mathcal{T}}_{x\to f}^{(1)}&=\lim_{\delta t\to 0+}\frac{1}{2\delta t}\log\frac{\sigma^{2}_{f(t+\delta t)|f(t)}}{\sigma^{2}_{f(t+\delta t)|f(t),x(t)}}\label{eqs:onestepxf},\\
\dot{\mathcal{T}}_{f\to x}^{(1)}&=\lim_{\delta t\to 0+}\frac{1}{2\delta t}\log\frac{\sigma^{2}_{x(t+\delta t)|x(t)}}{\sigma^{2}_{x(t+\delta t)|f(t),x(t)}}\label{eqs:onestepfx}.
\end{align}

The conditional variances can be derived from the equations of motion and the stationary covariance matrix. For the feedforward transfer entropy rate, Eq. \ref{eqs:spatialf} implies that $f(t+\delta t)$ is given by,
\begin{align}
f(t+\delta t)=f(t)-Ff(t)\delta t-Gkx(t)\delta t+\sqrt{2D_{f}\delta t}\mathcal{N}_{f},
\end{align}
where $\mathcal{N}_{f}$ is a standard normal random variable.
When both $f(t)$ and $x(t)$ are known, the variance in $f(t+\delta t)$ comes only from the noise, $\sigma_{f(t+\delta t)|f(t),x(t)}^{2}=2D_{f}\delta t$. When only $f(t)$ is known, the variance in $f(t+\delta t)$ is given by
\begin{align}
\sigma_{f(t+\delta t)|f(t)}^{2}=G^{2}k^{2}\delta t^{2}C_{x|f}+2D_{f}\delta t\label{eqs:condvarf1step},
\end{align}
where $C_{x|f}$ is the steady-state variance of $x$ conditioned on $f$. $C_{x|f}$ can be obtained from the stationary covariance matrix Eq. \ref{eqs:statcov} as
\begin{align}
C_{x|f}=C_{xx}-\frac{C_{xf}^{2}}{C_{ff}}=\frac{(F\beta+J\alpha)(F+H)\alpha-Gk\beta^{2}}{GkJ(F+H)\alpha\Delta},
\end{align}
where $\alpha=D_{v}Gk+D_{f}JH$, $\beta=D_{v}F(F+H)+D_{f}J^{2}$, and $\Delta=FH(F+H)-GkJ$.
Substituting these in Eq. \ref{eqs:onestepxf} gives,
\begin{align}
\dot{\mathcal{T}}_{x\to f}^{(1)}=\lim_{\delta t\to 0+}\frac{1}{2\delta t}\log\frac{G^{2}k^{2}\delta t^{2}C_{x|f}+2D_{f}\delta t}{2D_{f}\delta t}=\lim_{\delta t\to 0+}\frac{1}{2\delta t}\log\left[ 1+\frac{G^{2}k^{2}\delta t C_{x|f}}{2D_{f}}\right]&=\frac{G^{2}k^{2}C_{x|f}}{4D_{f}}\\
&=\frac{G^{2}k^{2}}{4D_{f}}\cdot \frac{(F\beta+J\alpha)(F+H)\alpha-Gk\beta^{2}}{GkJ(F+H)\alpha\Delta}\\
&=\frac{Gk(D_{v}F^{2}+D_{f}J^{2})}{4D_{f}FHJ}+\mathcal{O}(k^{2}).
\end{align}
The single-step feedforward transfer entropy can then be defined, analogous to the multi-step case, as
\begin{align}
\mathcal{T}_{\mathrm{FF}}^{(1)}\equiv\frac{\dot{\mathcal{T}}_{x\to f}^{(1)}}{H}&=\frac{Gk(D_{v}F^{2}+D_{f}J^{2})}{4D_{f}FH^{2}J}+\mathcal{O}(k^{2})\label{eqs:tffonestep1}\\
&=\frac{Gk}{2H^{2}}\sqrt{\frac{D_{v}}{D_{f}}}\cdot\frac{1}{2}\left[ \left(\frac{J}{F}\sqrt{\frac{D_{f}}{D_{v}}}\right)+\left(\frac{F}{J}\sqrt{\frac{D_{v}}{D_{f}}}\right) ^{-1}\right] \\
&=\mathcal{T}_{\mathrm{FF}}\cdot\frac{1}{2}\left[ \left(\frac{J}{F}\sqrt{\frac{D_{f}}{D_{v}}}\right)+\left(\frac{F}{J}\sqrt{\frac{D_{v}}{D_{f}}}\right) ^{-1}\right] ,
\end{align}
where, in the last step, we have substituted the expression for the multi-step feedforward transfer entropy from Eq. \ref{eqs:tff}. Through the AM-GM inequality,
\begin{align}
\mathcal{T}_{\mathrm{FF}}^{(1)}\geq \mathcal{T}_{\mathrm{FF}}.
\end{align}
This bound becomes tight at the optimal feedback strength, $J^{*}=F\sqrt{D_{v}/D_{f}}$. In the limits of weak and strong actuation strengths, $J\to0$ and $J\to\infty$, the single-step approximation makes a diverging error.

For deriving the single-step feedback transfer entropy rate Eq. \ref{eqs:onestepfx}, we can apply a similar procedure as above. From the equation of motion Eq. \ref{eqs:spatialx}, $x(t+\delta t)$ is given by,
\begin{align}
x(t+\delta t)=x(t)+v(t)\delta t.
\end{align}
When only $x(t)$ is known, the variance in $x(t+\delta t)$ arises from the conditional variance of $v(t)$, $\sigma^{2}_{x(t+\delta t)|x(t)}=\delta t^{2}C_{v|x}$. When both $x(t)$ and $f(t)$ are known, the variance in $x(t+\delta t)$ stems from a different conditional variance, $\sigma^{2}_{x(t+\delta t)|f(t),x(t)}=\delta t^{2}C_{v|xf}$. Strikingly, this implies that the transfer entropy rate diverges,
\begin{align}
\dot{\mathcal{T}}_{f\to x}^{(1)}=\lim_{\delta t\to 0+}\frac{1}{2\delta t}\log\frac{C_{v|x}}{C_{v|fx}}\to\infty,
\end{align}
since the steady-state variances $C_{v|x}$ and $C_{v|fx}$ are both finite and independent of $\delta t$.
Therefore, the single-step approximation to the feedback transfer entropy rate makes a diverging error.

In order to obtain physically meaningful single-step transfer entropy rates, we next compute the transfer entropy between $v$ and $f$.

\subsection*{Single-step transfer entropy between $v$ and $f$}
Under the single-step approximation, the output of the actuation system, $v(t)$, does not map one-to-one to the sensory input $x(t)$. Therefore, the information transmission through the actuator could alternatively be defined as the single-step transfer entropy from $f(t)$ to $v(t)$. Here, we compute this alternative. For the sake of completion, we also compute the single-step transfer entropy from $v(t)$ to $f(t)$. However, for all definitions of the single-step feedforward and feedback transfer entropy, a BEST remains elusive.

The single-step transfer entropy rates between $v$ and $f$ are defined as
\begin{align}
\dot{T}_{v\to f}^{(1)}&=\lim_{\delta t\to0+}\frac{1}{\delta t}\left\langle\log\frac{P[f(t+\delta t)|f(t),v(t)]}{P[f(t+\delta t)|f(t)]}\right\rangle ,\\
\dot{T}_{f\to v}^{(1)}&=\lim_{\delta t\to0+}\frac{1}{\delta t}\left\langle\log\frac{P[v(t+\delta t)|f(t),v(t)]}{P[v(t+\delta t)|v(t)]}\right\rangle .
\end{align}
Since the dynamics is linear, the transfer entropy rates are given by conditional variances,
\begin{align}
\dot{\mathcal{T}}_{v\to f}^{(1)}&=\lim_{\delta t\to 0+}\frac{1}{2\delta t}\log\frac{\sigma^{2}_{f(t+\delta t)|f(t)}}{\sigma^{2}_{f(t+\delta t)|f(t),v(t)}},\\
\dot{\mathcal{T}}_{f\to v}^{(1)}&=\lim_{\delta t\to 0+}\frac{1}{2\delta t}\log\frac{\sigma^{2}_{v(t+\delta t)|v(t)}}{\sigma^{2}_{v(t+\delta t)|f(t),v(t)}}.
\end{align}
Analogous to the previous section, we compute the conditional covariances from the equations of motion, Eqs. \ref{eqs:spatialx}-\ref{eqs:spatialv}, and the stationary covariance matrix in Eq. \ref{eqs:statcov}. The conditional variance $\sigma^{2}_{f(t+\delta t)|f(t)}$ was shown, in Eq. \ref{eqs:condvarf1step}, to be $G^{2}k^{2}\delta t^{2}C_{x|f}+2D_{f}\delta t$. When both $f(t)$ and $v(t)$ are known, the conditional variance of $f(t+\delta t)$ is $\sigma^{2}_{f(t+\delta t)|f(t),v(t)}=G^{2}k^{2}\delta t^{2}C_{x|fv}+2D_{f}\delta t$. This gives the single-step feedforward transfer entropy rate,
\begin{align}
\dot{\mathcal{T}}_{v\to f}^{(1)}=\lim_{\delta t\to 0+}\frac{1}{2\delta t}\log\frac{G^{2}k^{2}\delta t^{2}C_{x|f}+2D_{f}\delta t}{G^{2}k^{2}\delta t^{2}C_{x|fv}+2D_{f}\delta t}=\lim_{\delta t\to 0+}\frac{1}{2\delta t}\log\frac{1+\frac{G^{2}k^{2}\delta t C_{x|f}}{2D_{f}}}{1+\frac{G^{2}k^{2}\delta t C_{x|fv}}{2D_{f}}}&=\frac{G^{2}k^{2}}{4D_{f}}(C_{x|f}-C_{x|fv}).
\end{align}
We can obtain this from the stationary covariance matrix as 
\begin{align}
C_{x|f}-C_{x|fv}=C_{xx}-\frac{C^{2}_{xf}}{C_{ff}}-C_{xx}+\mathbf{C}_{x,fv}\mathbf{C}_{fv,fv}^{-1}\mathbf{C}_{x,fv}^{T}=\mathbf{C}_{x,fv}\mathbf{C}_{fv,fv}^{-1}\mathbf{C}_{x,fv}^{T}-\frac{C^{2}_{xf}}{C_{ff}},
\end{align}
where $\mathbf{C}_{x,fv}$ and $\mathbf{C}_{fv,fv}$ are sub-matrices of the full covariance matrix $\mathbf{C}$ in Eq. \ref{eqs:statcov}. This gives, after simplification,
\begin{align}
\dot{\mathcal{T}}_{v\to f}^{(1)}=\frac{G^{2}k^{2}}{4D_{f}}\frac{\beta^{2}}{(F+H)(F\beta+D_{v}\Delta)\Delta}=\frac{G^{2}k^{2}[D_{v}F(F+H)+D_{f}J^{2}]^{2}}{4D_{f}F^{2}H(F+H)^{2}[D_{v}(F+H)^{2}+D_{f}J^{2}]}+\mathcal{O}(k^{3}),
\end{align}
where $\beta=D_{v}F(F+H)+D_{f}J^{2}$ and $\Delta=FH(F+H)-GkJ$, as before. Therefore, the corresponding feedforward transfer entropy is
\begin{align}
\mathcal{T}_{\mathrm{FF}}^{(1)'}=\frac{\dot{\mathcal{T}}_{v\to f}^{(1)}}{H}=\frac{G^{2}k^{2}[D_{v}F(F+H)+D_{f}J^{2}]^{2}}{4D_{f}F^{2}H^{2}(F+H)^{2}[D_{v}(F+H)^{2}+D_{f}J^{2}]}+\mathcal{O}(k^{3})\label{eqs:tffonestep2}.
\end{align}

Analogously, we derive the alternative single-step feedback transfer entropy rate using the equation of motion for $v(t)$, Eq. \ref{eqs:spatialv}. The temporal update of $v(t)$ is
\begin{align}
v(t+\delta t)=v(t)-Hv(t)\delta t+Jf(t)\delta t+\sqrt{2D_{v}\delta t}\mathcal{N}'_{v},
\end{align}
where $\mathcal{N}'_{v}$ is a standard normal random variable. When both $f(t)$ and $v(t)$ are known, the variance in $v(t+\delta t)$ stems solely from the noise, $\sigma^{2}_{v(t+\delta t)|f(t),v(t)}=2D_{v}\delta t$. When only $v(t)$ is known, the conditional variance is $\sigma^{2}_{v(t+\delta t)|v(t)}=J^{2}\delta t^{2}C_{f|v}+2D_{v}\delta t$. The conditional variance $C_{f|v}$ is obtained from Eq. \ref{eqs:statcov} as $C_{f|v}=C_{ff}-C^{2}_{fv}/C_{vv}$. This gives the single-step feedback transfer entropy rate as
\begin{align}
\dot{\mathcal{T}}_{f\to v}^{(1)}=\lim_{\delta t\to 0+}\frac{1}{2\delta t}\log\frac{J^{2}\delta t^{2}C_{f|v}+2D_{v}\delta t}{2D_{v}\delta t}=\lim_{\delta t\to 0+}\frac{1}{2\delta t}\log\left[ 1+\frac{J^{2}\delta t C_{f|v}}{2D_{v}}\right]&=\frac{J^{2}C_{f|v}}{4D_{v}}\\
&=\frac{J^{2}}{4D_{v}}\cdot \frac{\alpha(F\beta+D_{v}\Delta)}{J\beta\Delta}\\
&=\frac{J^{2}D_{f}[D_{v}(F+H)^{2}+D_{f}J^{2}]}{4D_{v}(F+H)[D_{v}F(F+H)+D_{f}J^{2}]}+\mathcal{O}(k),
\end{align}
where $\alpha=D_{v}Gk+D_{f}JH$, $\beta=D_{v}F(F+H)+D_{f}J^{2}$, and $\Delta=FH(F+H)-GkJ$, as before. So, the single-step feedback transfer entropy is
\begin{align}
\mathcal{T}_{\mathrm{FB}}^{(1)'}=\frac{\dot{\mathcal{T}}_{f\to v}^{(1)}}{F}=\frac{J^{2}D_{f}[D_{v}F(F+H)^{2}+D_{f}J^{2}]}{4D_{v}(F+H)[D_{v}F(F+H)+D_{f}J^{2}]}+\mathcal{O}(k)\label{eqs:tfbonestep2}.
\end{align}

The existence of a BEST equation requires the navigation performance to be expressed through the feedforward and feedback transfer entropy expressions without fitting or scaling parameters. The navigation performance is given in Eq. \ref{eqs:fullperf} as
\begin{align}
\frac{\mathcal{P}}{\mathcal{P}_{0}}=\frac{GkJD_{v}F}{H^{2}(D_{f}J^{2}+D_{v}F^{2})}+\mathcal{O}(k^{2})
\end{align}
From the expressions of the two alternate definitions of the single-step feedforward transfer entropy, $\mathcal{T}_{\mathrm{FF}}^{(1)}$ and $\mathcal{T}_{\mathrm{FF}}^{(1)'}$ in Eqs. \ref{eqs:tffonestep1} and \ref{eqs:tffonestep2}, and the expression for the single-step feedback transfer entropy, $\mathcal{T}_{\mathrm{FB}}^{(1)'}$ in Eq. \ref{eqs:tfbonestep2}, we find that a BEST equation cannot be derived. Therefore, single-step transfer entropies are not the key collective variables that govern navigation.